\documentclass[10pt,journal,cspaper,compsoc]{IEEEtran}

\pdfoutput=1

\RequirePackage{ifpdf}
\ifpdf
  \usepackage[pdftex,breaklinks,hyperfootnotes=false,bookmarks=false]{hyperref}
\else
  \usepackage[hyperfootnotes=false,bookmarks=false]{hyperref}
\fi
\hypersetup{
  colorlinks=true,linkcolor=black,citecolor=black,urlcolor=black,breaklinks=true
}

\RequirePackage{graphicx}

\usepackage{epstopdf}

\usepackage[cmex10]{amsmath}
\usepackage{array}
\usepackage{mdwmath}
\usepackage{mdwtab}
\usepackage{cite}
\usepackage{amsmath,amssymb,amsthm}

\usepackage{eqparbox}

\usepackage{enumerate}

\usepackage{epstopdf}

\ifCLASSOPTIONcompsoc
    \usepackage[tight,normalsize,sf,SF]{subfigure}
\else
    \usepackage[tight,footnotesize]{subfigure}
\fi

\begin{document}
%
\title{A Closed-Form Formulation of HRBF-Based Surface Reconstruction}
%
%
%
%
\author{Shengjun~Liu,
        Charlie~C.~L.~Wang,~\IEEEmembership{Senior Member,~IEEE,}
        Guido~Brunnett,
        and~Jun~Wang
\IEEEcompsocitemizethanks{\IEEEcompsocthanksitem S. Liu is with the School of Mathematics and Statistics, Central South University, Changsha 410083, China.\protect\\
E-mail: shjliu.cg@gmail.com
\IEEEcompsocthanksitem C.C.L. Wang is with Department of Mechanical and Automation Engineering, The Chinese University
of Hong Kong, Shatin, N.T., Hong Kong. Corresponding author - E-mail: cwang@mae.cuhk.edu.hk
\IEEEcompsocthanksitem G. Brunnett is with Department of Computer Science,
Chemnitz University of Technology, Chemnitz 09126, Germany.
\IEEEcompsocthanksitem J. Wang is with the College of
Mechanical and Electrical Engineering, Nanjing University of Aeronautics and Astronautics, Nanjing 210016, China.}
\thanks{}}
%
%

\markboth{Manuscript Submitted}%
{HRBF}
%


\IEEEcompsoctitleabstractindextext{%
\begin{abstract}
The \textit{Hermite radial basis functions} (HRBFs) implicits have been used to reconstruct surfaces from scattered
Hermite data points. In this work, we propose a closed-form formulation to construct HRBF-based implicits by a
quasi-solution approximating the exact solution. A scheme is developed to automatically adjust the support sizes of
basis functions to hold the error bound of a quasi-solution. Our method can generate an implicit function from
positions and normals of scattered points without taking any global operation. Working together with an adaptive
sampling algorithm, the HRBF-based implicits can also reconstruct surfaces from point clouds with non-uniformity and
noises. Robust and efficient reconstruction has been observed in our experimental tests on real data captured from a
variety of scenes.
\end{abstract}

}

\maketitle

\IEEEdisplaynotcompsoctitleabstractindextext

 \ifCLASSOPTIONpeerreview
 \begin{center} \bfseries EDICS Category: 3-BBND \end{center}
 \fi
%
\IEEEpeerreviewmaketitle

\section{Introduction}
\IEEEPARstart{R}{econstructing} surface from a set of unorganized points equipped with normal vectors is an important
topic in various fields such as computer graphics, reverse engineering, image processing, mathematics, robotics and
CAD/CAM. A lot of research approaches have been devoted to develop surface reconstruction methods, in which implicit
surface fitting based on \textit{Radial Basis Functions} (RBF) is successful in dealing with noisy and incomplete data
(e.g., \cite{Carr:2001,Turk:2002,Wendland:2005}).

Recently, implicits based on \textit{Hermite Radial Basis Functions} (HRBF) were presented to interpolate data points
to the first order~\cite{Macedo:2011}. It is robust and effective to deal with coarse and non-uniformly sampled points,
close surface sheets, and surfaces with fine details. However, interpolating both positions and normals of points leads
to the computation of solving a $4n \times 4n$ linear system for an input with $n$ points. It is impractical due to the
expensive computation. The system becomes sparse when the \textit{Compactly Supported Radial Basis Functions} (CSRBF)
are used as the kernel functions. However, this attempt on improving the efficiency can bring in a more challenging
problem -- numerical stability.
\begin{figure}
\includegraphics[width=\linewidth]{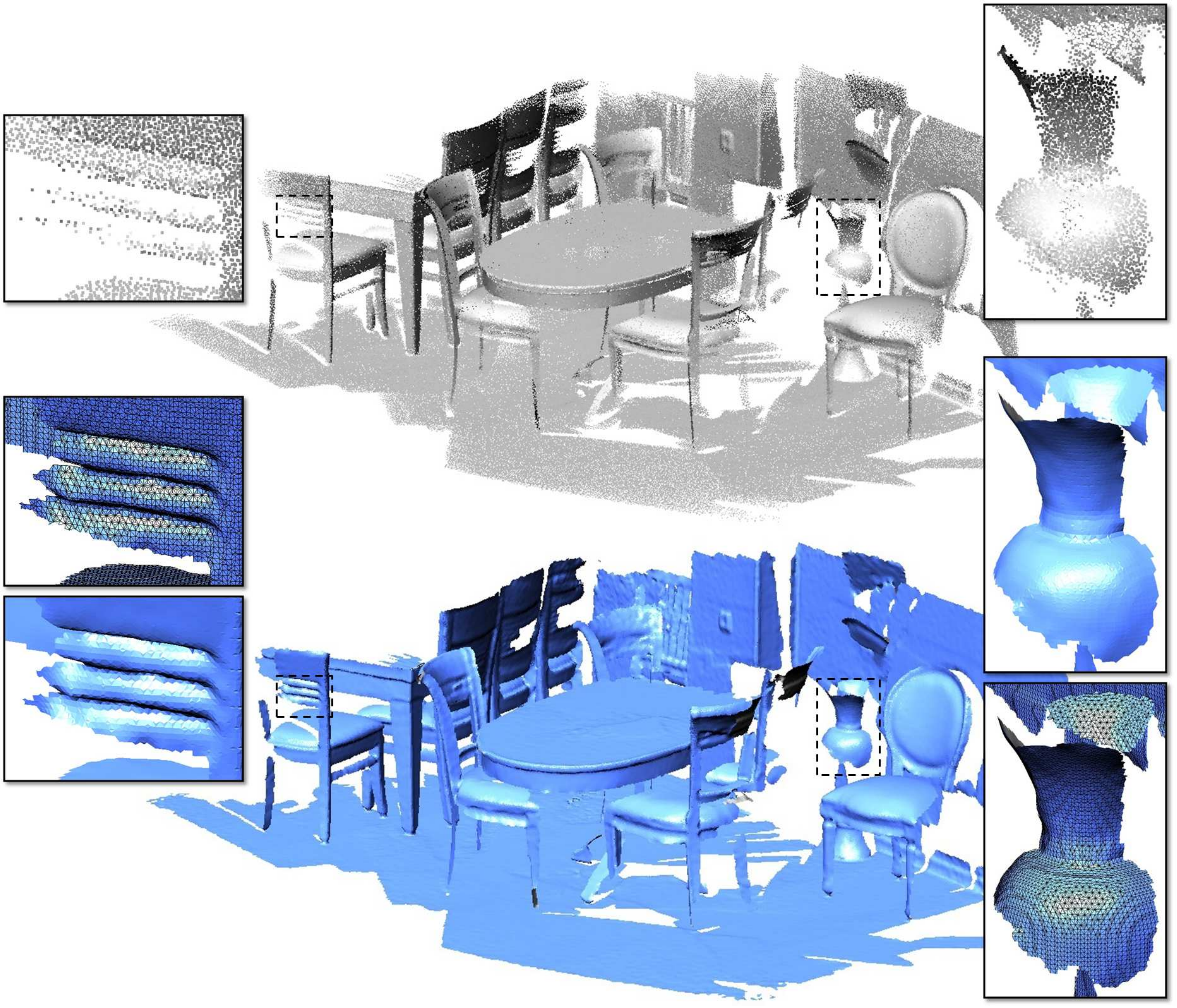}
\caption{The method proposed in this paper can efficiently reconstruct a surface from a set of noisy and incomplete
points -- e.g., the indoor scene shown here with 922k points. Our reconstruction takes only 60 seconds to generate a
mesh surface that has the similar quality as the state-of-the-art~\cite{Fuhrmann:2014} but the computation is $7.6
\times$ faster. The parameter, $s=3.5$, is employed in this reconstruction.}\label{fig:IndoorReconstructionExample}
\end{figure}

HRBF-based quasi-interpolation is presented in this paper to overcome the computational problem. Quasi-interpolation is
a kind of approximate interpolation that fits implicits by weighted averages of the values at given points. The most
attractive property of quasi-interpolation is to reconstruct a surface from a set of points without solving linear
systems -- i.e., with a closed-form formulation. This can make the computation of HRBF-based surface reconstruction
stable and efficient. As shown in Fig.\ref{fig:IndoorReconstructionExample}, the mesh surface can be efficiently
reconstructed from an input set with $922k$ points by our method in $60$ seconds. Comparing to the recently published
\textit{Floating Scale Surface Reconstruction} (FSSR) that also avoids applying global operations, our method is about
$7.6 \times$ faster. $s=3.5$ is the \textit{only} parameter used in our approach. Here, $s$ is defined as an amplifier
of CSRBF kernels' support sizes, which control the maximal number of data points covered by each support (details can
be found in Section \ref{subsecSupportSize}). Moreover, we have analyzed the error-bound between our closed-form
solution and the solution obtained by solving linear systems (see Section \ref{subsecErrorBound}). Specifically, the
error-bound exists when the number of points covered by the support of each kernel is capped by a fixed number. To
overcome the problem caused by high non-uniformity on points, an algorithm is developed to select a sub-set of given
points to serve as centers of kernel functions, which is an optional step in our framework of reconstruction.

\subsection{Main result}
In this paper, we propose a closed-form formulation for computing the quasi-solution of HRBF-based surface
reconstruction from scattered data points.
\begin{itemize}
\item The method can construct a signed scalar function by directly blending the positions and normals of points
without any global operation. The computation based on CSRBF is local and robust.

\item Errors between the quasi-solution and the exact one are bounded after applying an automatical scheme to adjust
the support sizes of basis functions.

\item Combining with an adaptive center selection algorithm, surface reconstruction based on our method can remove the
artifacts resulted from the non-uniformity and noises.
\end{itemize}
As a local approach, our method is efficient and scalable. This is well-suited for highly parallel implementation as
well as distributed/progressive reconstruction. Note that, the compactly-supported basis functions results in open
meshes and leaves holes in the region does not have enough number of points, which fits the application of
reconstructing outdoor scenes very well.

\subsection{Related work}
The problem of surface reconstruction from point cloud has been studied in literature for more than two decades. A
comprehensive review of all these works has beyond the scope of our paper. More discussion and comparison on different
surface reconstruction methods can be found in \cite{Berger:2013}. We only give an overview of implicit function based
reconstruction with a focus on RBF-based formulations.

After using signed distance field in~\cite{Hoppe:1992} to reconstruct mesh surface from point clouds, implicit
functions have gained a lot of attention in surface reconstruction because of its ability to handle topological change
and fill holes. Example approaches include RBF-based
methods~\cite{Macedo:2011,Carr:2001,Turk:2002,Morse:2001,Ohtake:2005,Pan:2009,Samozino:2006,Sussmuth:2010,Ohtake:2006,Tobor:2006,Walder:2006,Alliez:2007},
Poisson surface reconstructions~\cite{Kazhdan:2006,Kazhdan:2013}, smooth signed distance method~\cite{Calakli:2011},
moving least-squares based methods~\cite{Alexa:2001,Guennebaud:2007,Amenta:2004,Oztireli:2009,Alexa:2009,Yang:2009},
wavelets based method~\cite{Manson:2008}, and \textit{Partition-Of-Unity} (POU) based
methods~\cite{Gois:2008,Ohtake:2003}. The methods based on RBF implicits are popular for their capability of handling
sparse point clouds. Generally, RBF-based methods transform the reconstruction into a problem of multi-variational
optimization, where enforcing the interpolation constraints results in a linear system. Solving the linear system is an
important but time-consuming step for the RBF-based reconstruction. To obtain a non-trivial solution, RBF-based methods
usually require the provision of extra offset-points (ref. \cite{Carr:2001,Turk:2002}) that can be obtained by shifting
data points along their normal directions. However, it is not easy to find an optimal value for the offset scalar. The
positions of these offset points is also difficult to determine, especially when the scanned model has thin sheets and
the distribution of input points is irregular. To avoid generating offset-points, Ohtake et al.
\cite{Ohtake:2005,Ohtake:2006} used a signed function which includes basic approximations and local details. The basic
approximation formed by local quadratic functions with POU is considered as an offset function which constructs the
non-zero constraints for fitting local details with RBFs. A very important information that describes shape to the
first order -- the normal vectors of a model have not been well utilized in these approaches of surface reconstruction.

Some prior works (e.g.,~\cite{Pan:2009,Walder:2006}) deduced from the statistical-learning perspective avoid generating
offset-points in surface reconstruction, where normals were directly used in a variational formulation. Recently,
Macedo et al. \cite{Macedo:2011} derive an implicit function from the Hermite-Birkhoff interpolation with RBFs. They
enhance the flexibility of HRBF reconstruction by ensuring well-posedness of an approach combining regularization.
However, given a set with $n$ points and $n$ normal vectors, these methods give a $4n\times 4n$ linear system to be
solved, which limits the number of points can be involved in the reconstruction.

Quasi-interpolation is a method in the field of function approximation. It is simple, efficient, and computational
stable. In the early work of quasi-interpolation~\cite{Shepard:1968}, a function approximating a given data set is
defined by a weighted average of the values at the data points. The idea has been used in \cite{Xie:2004} for surface
reconstruction. The quasi-interpolation with radial basis functions has been studied in \cite{Buhmann:1993}. Recently,
Wu and Xiong \cite{Wu:2010} developed a new method to construct kernels in a quasi-interpolation scheme by the linear
combination of scales, where the kernels are in the form of RBF. Han and Hou discussed quasi-interpolation by RBF and
suggested values of the shape parameters in \cite{Han:2008}, which is a constructive method for obtaining a family of
quasi-interpolations. Liu and Wang generalized the regularly sampled grid points to 3D scattered points in
\cite{Liu:2012}. They proposed a multi-level quasi-interpolation method which is based on POU and the RBF method.
However, linear systems still need to be solved in their method. Locally supported basis functions satisfying the
property of POU are used in the recent work FSSR~\cite{Fuhrmann:2014}, in which weighted average is employed to fit
implicit functions to the input set of points. However, there is no error-bound guaranteed in their approach.
Experimental tests show that our method can generate results with similar quality as FSSR but has $7.61$ to $98.3$
times speedup.

There are schemes for finding a subset of `optimal' centers from the point set in order to obtain fast reconstructions.
Carr et al. \cite{Carr:2001} proposed a greedy algorithm that iteratively appends centers which are corresponding to
the maximal residual of the current RBF fitting until a desired accuracy is reached. Samozino \textit{et al.} presented
the reconstruction with voronoi centered RBFs \cite{Samozino:2006}. Ohtake et al. \cite{Ohtake:2006} proposed a
reconstruction method which combines an adaptive POU approximation with least-squares RBF fitting. Different
from~\cite{Ohtake:2006}, we adopt a quadric error function based on positions and normals of a point and its neighbors
instead of local quadratic approximation. Therefore, our center selection step can generate a good spherical cover in a
non-iterative way. The selected centers of spheres describe the input shape with a small error.

\vspace{8pt} The rest of our paper is organized as follows. We first introduce the surface reconstruction based on
regularized HRBF implicits in Section \ref{secHRBFImplicits}. Section \ref{secFormulation} provides our formulation in
the closed-form and derives the error bound of our formulation. Section \ref{secReconstruction} presents the algorithms
for different steps of reconstruction, including parameters tuning, isosurface extraction and center selection
(optional). After that, the results of experimental tests are shown and discussed in Section \ref{secResult}. Lastly,
our paper ends with the conclusion.

\section{HRBF Implicits}\label{secHRBFImplicits}
The HRBF implicits \cite{Macedo:2011} are built upon the theory of Hermite-Birkhoff interpolation with radial basis
functions \cite{Wendland:2005}. In this section, we briefly describe how to use HRBF implicits to solve the problem of
surface reconstruction from scattered points.

\vspace{8pt} \noindent \textbf{Definition 1}~~Given a set of data $\mathcal{P}=\{\mathbf{p}_1, \mathbf{p}_2, \cdots,
\mathbf{p}_n\}$ with unit normals $\mathcal{N}=\{\mathbf{n}_1, \mathbf{n}_2, \cdots, \mathbf{n}_n\}$, the HRBF
implicits give a function $f$ interpolating both the points and the normal vectors as
\begin{equation}\label{eq:eqHRBFInterpolation}
f(\mathbf{x})=\sum_{j=1}^{n}{ \{ a_j \varphi(\mathbf{x}-\mathbf{p}_j)- \langle \mathbf{b}_j,
\nabla{\varphi(\mathbf{x}-\mathbf{p}_j)} \rangle \}},
\end{equation}
where $\varphi:\Re^3 \mapsto \Re$ is defined by a radial basis function
$\varphi(\mathbf{x})=\phi_{\rho}(\|\mathbf{x}\|)$, $\langle\cdot,\cdot\rangle$ denotes the dot-product of two vectors,
and $\nabla$ is the gradient operator. \vspace{8pt}

The scalar coefficients, $a_j\in\Re$, and the vector coefficients, $\mathbf{b}_j\in\Re^3$, can be determined by the
constraints of interpolation as
\begin{equation}\label{eq:eqInterpolationConstraints}
f(\mathbf{p}_i)=c \;\; \mathrm{and} \;\; \nabla{f(\mathbf{p}_i)}=\mathbf{n}_i, \;\; (i=1,2,\cdots,n)
\end{equation}
with $c$ being a constant value for the implicit function. $c=0$ is used for surface reconstruction. Applying the
constraints (\ref{eq:eqInterpolationConstraints}) to Eq.(\ref{eq:eqHRBFInterpolation}), we obtain a linear system with
equations
\begin{equation}\label{eq:eqInterLinearSystem}
\left.%
\begin{array}{l}
\sum_{j=1}^{n}{\{a_j\varphi(\mathbf{p}_i-\mathbf{p}_j)-\langle \mathbf{b}_j, \nabla{\varphi(\mathbf{p}_i-\mathbf{p}_j)}\rangle\}}=c, \\
\sum_{j=1}^{n}{\{a_j\nabla\varphi(\mathbf{p}_i-\mathbf{p}_j)-\mathbf{b}_j \mathbf{H}\varphi(\mathbf{p}_i-\mathbf{p}_j)\} }=\mathbf{n}_i, \\
\end{array}%
\right.
\end{equation}
where $i=1,2,\cdots,n$ and $\mathbf{H}$ is the Hessian operator applied on $\varphi(\cdot)$.
The linear system can be rewritten in a matrix form as
\begin{equation}\label{eq:eqInterLinearSystemMatrix}
\mathbf{A}\boldsymbol{\lambda}=\mathbf{y},
\end{equation}
where $\boldsymbol{\lambda}$ and $\mathbf{y}$ are $4n$ vectors
with the $i$-th blocks being $[a_i, \mathbf{b}_i]^T$ and
$[c,\mathbf{n}_i]^T$ respectively. Here, $\mathbf{A}$ is a $4n
\times 4n$ coefficients matrix which are assembled from $n \times
n$ blocks. Each block $\mathbf{A}_{i,j}$ is a $4\times 4$
sub-matrix corresponding to a pair of RBF centers
$(\mathbf{p}_i,\mathbf{p}_j)$.
\begin{equation}\label{eq:coefficientMatrixBlocks}
\begin{aligned}
&\mathbf{A}=(\mathbf{A}_{i,j})_{n\times n},\\
&\mathbf{A}_{i,j}= {\begin{pmatrix}
\varphi(\mathbf{p}_i-\mathbf{p}_j) & -\nabla\varphi(\mathbf{p}_i-\mathbf{p}_j)\\
\nabla\varphi(\mathbf{p}_i-\mathbf{p}_j) &
-\mathbf{H}\varphi(\mathbf{p}_i-\mathbf{p}_j)
\end{pmatrix}}_{4\times 4}.
\end{aligned}
\end{equation}

\begin{figure}[t]\small
\centering
\includegraphics[width=0.45\linewidth]{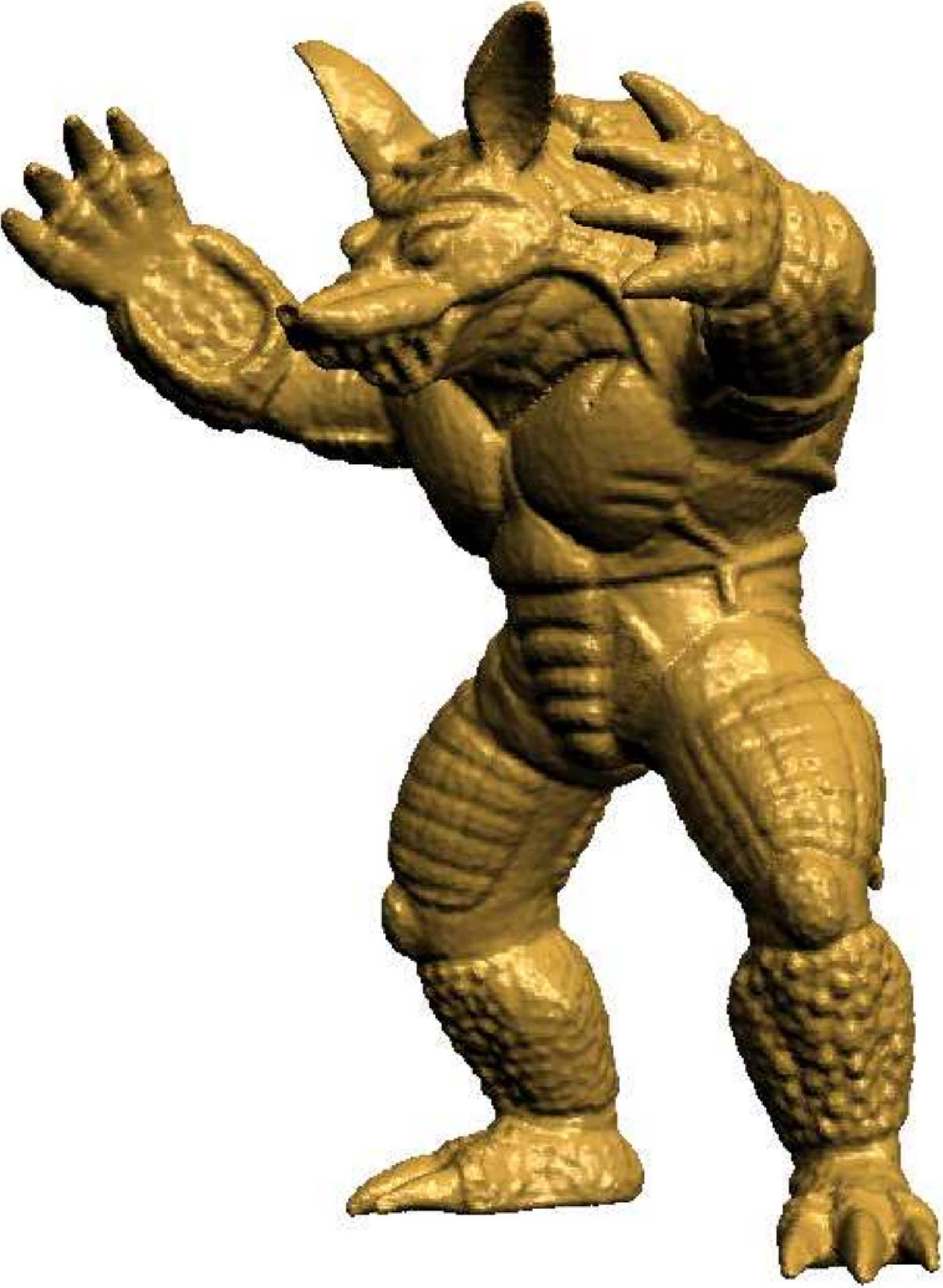}
\includegraphics[width=0.45\linewidth]{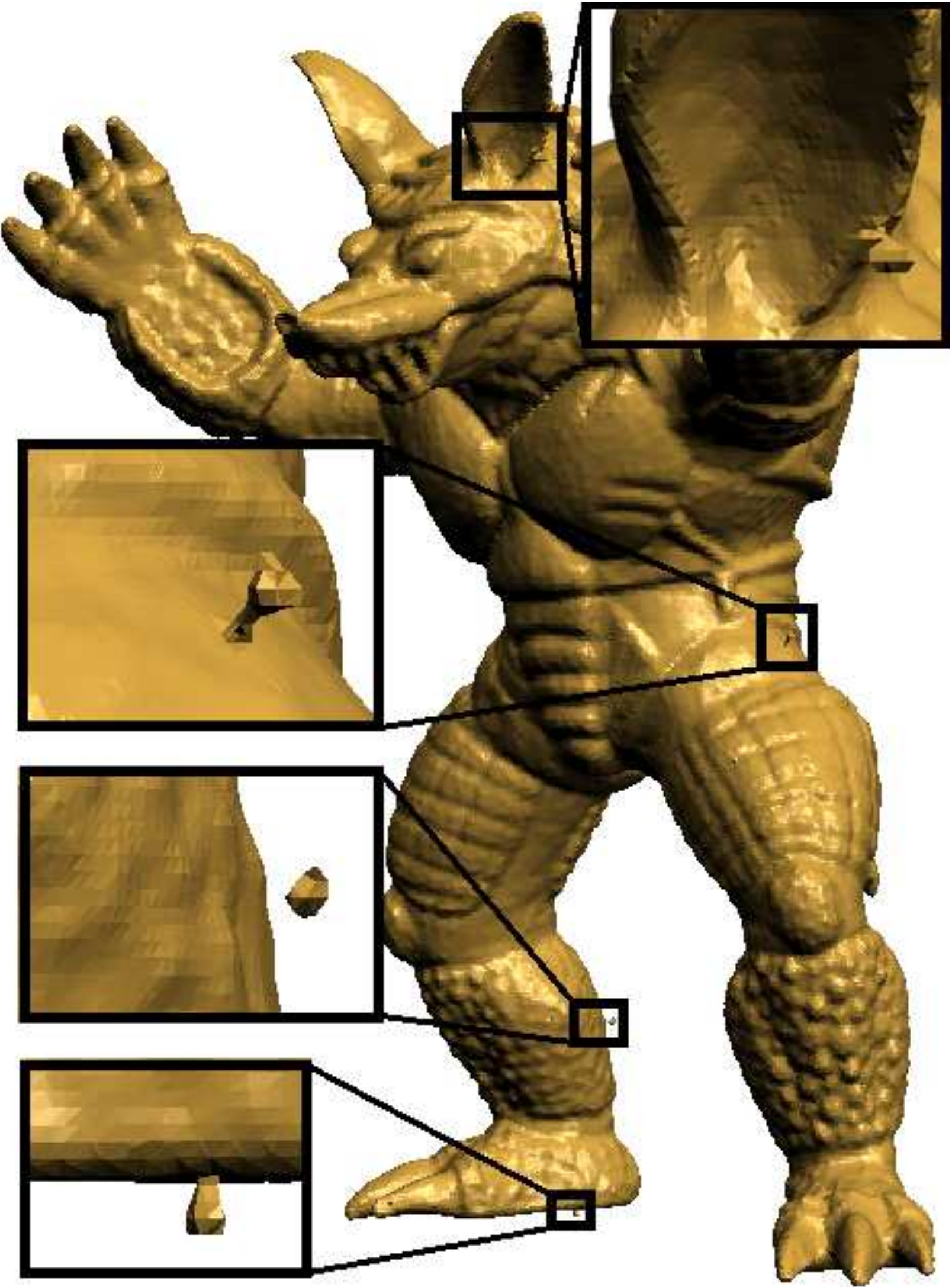}
\caption{Surface reconstructions by the HRBF implicits with (left) and without (right) the regularization term.
Artifacts will be produced when the regularization term is \textit{not} added. In this example, $\eta=100/\rho^2$ is
employed.}\label{fig:Regularization}
\end{figure}

In this paper, we use a Wendaland's CSRBF \cite{Wendland:1995} as
the kernel function
\begin{equation}\label{eq:CSRBF}
\begin{aligned}
&  \phi_{\rho}(r) = \phi(r/\rho) \\
&  \phi(t) = \begin{cases}
(1-t)^4(4t+1),  & t\in [0,1], \\
0,  & \text{otherwise},
\end{cases}
\end{aligned}
\end{equation}
where $\rho$ is the support size, and $r$ is the Euclidean distance between a query point and the center of a RBF. Note
that, different support sizes can be used at different centers. Solving Eq.(\ref{eq:eqInterLinearSystemMatrix}), an
implicit function $f(\mathbf{x})$ can be determined at the space spanned by the supports of centers $\{\mathbf{p}_i\}$.
To make the system matrix of RBF interpolation better conditioned, a regularization term with coefficient $\eta$ is
usually added when using RBF interpolants to solve the surface reconstruction problem (ref.~\cite{Dinh:2002}). That is,
\begin{equation}\label{eq:eqRegularizedLinearSystemMatrix}
(\mathbf{A}+ \eta \mathbf{I}) \boldsymbol{\lambda}=\mathbf{y}.
\end{equation}
An example is shown in Fig.\ref{fig:Regularization} to demonstrate the effectiveness of regularization.

\section{Formulation}\label{secFormulation}
This section provides a closed-form formulation for solving the HRBF-based surface reconstruction problem via
quasi-solution. Error-bound of the approximation is also derived.

\subsection{Quasi-solution in closed-form}\label{subsecQuasiSolution}
When increasing the number of centers in the linear system of HRBF-based surface reconstruction (i.e.,
Eq.(\ref{eq:eqRegularizedLinearSystemMatrix})), numerical instability and expensive computation become intensively
remarkable. Here, we investigate a closed-form formulation derived from the theorem of quasi-interpolation to
reconstruct surface in a more stable and efficient way.

Quasi-interpolation technique can reconstruct a function interpolating a given data set by computing weighted averages
of the values at the data points \cite{Shepard:1968}. Specifically, considering an exact interpolant
$g(\mathbf{x})=\sum_{i}\lambda_i\psi_i(\mathbf{x})$ with the constraints $g(\mathbf{x}_i)=f_i$ of function values, the
function $g(\mathbf{x})$ can be well approximated by letting $\lambda_i \equiv f_i$. That is a quasi-interpolation,
$\tilde{g}(\mathbf{x})=\sum_{i}f_i\psi_i(\mathbf{x})$. However, the quasi-interpolation technique cannot be directly
applied here as our interpolation constraints consist of both the values and the gradients of functions (see
Eq.(\ref{eq:eqInterpolationConstraints})).

Basically, we need a closed-form formulation to approximate the solution of
Eq.(\ref{eq:eqRegularizedLinearSystemMatrix}). By means of the matrix computation, the quasi-interpolant with
$\lambda_i \equiv f_i$ is actually a quasi-solution when the coefficient matrix is approximated by an identity matrix
$\mathbf{I}$. Here, a similar approximation is employed in the HRBF-based reconstruction problem. For a CSRBF
$\varphi_i(\cdots)$, when there is no other center falling into the space spanned by its support $\rho_i$, the
coefficient matrix is degenerated from $\mathbf{A}_{i,i}$ of Eq.(\ref{eq:coefficientMatrixBlocks}) into
\begin{equation}\label{eq:D_matrix}
\mathbf{D}_{i,i}= diag (1, \frac{20}{\rho_i^2}, \frac{20}{\rho_i^2}, \frac{20}{\rho_i^2})  + \eta \mathbf{I}_4, \quad
\mathbf{D}_{i,j}=0 \; (i\neq j).
\end{equation}
If the scenario of not containing other centers happens at all CSRBF kernels, the linear system to be solved in
Eq.(\ref{eq:eqRegularizedLinearSystemMatrix}) is degenerated into $\mathbf{D} \boldsymbol{\lambda}=\mathbf{y}$ with
$\mathbf{D}=(\mathbf{D}_{i,j})_{n \times n}$. This leads to an approximate solution of
Eq.(\ref{eq:eqRegularizedLinearSystemMatrix}) as
\begin{equation}\label{eq:quasiSolution}
\begin{array}{rl}
  \tilde{\boldsymbol{\lambda}} & = \mathbf{D}^{-1} \mathbf{y} \\
   & = \{ \frac{c}{1+\eta},\frac{ \rho_1^2 \mathbf{n}_1}{20+\eta
\rho_1^2}, \cdots, \frac{c}{1+\eta},\frac{ \rho_n^2
\mathbf{n}_n}{20+\eta \rho_n^2} \}. \\
\end{array}
\end{equation}
The zero level-set is usually employed in surface reconstruction (i.e., $c=0$ in
Eqs.(\ref{eq:eqInterpolationConstraints}) and (\ref{eq:eqInterLinearSystem})). $c=0$ is used in all formulas in the
rest of this paper. As a result, the coefficients of the $i$-th basis function can be approximated by
\[
a_i \approx 0 \quad \mathrm{and} \quad \mathbf{b}_i \approx \frac{
\rho_i^2 }{20+\eta \rho_i^2}\mathbf{n}_i,
\]
which give an approximate function of $f(\mathbf{x})$ in a closed form as
\begin{equation}\label{eq:appFuncClosedForm}
\tilde{f}(\mathbf{x})=-\sum_{j=1}^{n}\langle\frac{\rho_j^2}{20+\eta
\rho_j^2}\mathbf{n}_j,\nabla\varphi(\mathbf{x}-\mathbf{p}_j)\rangle.
\end{equation}
By this implicit function, we can apply the polygonization techniques (e.g., \cite{Lorensen:1987,Ju:2002}) to
tessellate the isosurface of $\tilde{f}(\mathbf{x})=0$ into a polygonal mesh -- the surface is reconstructed from
scattered Hermite points.

\subsection{Error-bound analysis}\label{subsecErrorBound}
The error between the quasi-solution
$\tilde{\boldsymbol{\lambda}}$ and the exact solution
$\boldsymbol{\lambda}$ of
Eq.(\ref{eq:eqRegularizedLinearSystemMatrix}) must be bounded to
make the closed-form formulation useful. The analysis is given
below.

\vspace{8pt} \noindent \textbf{Lemma 1} \hspace{5pt} Defining
$\Delta\mathbf{A}= (\mathbf{A} + \eta \mathbf{I}) - \mathbf{D}$
and
$\Delta\boldsymbol{\lambda}=\boldsymbol{\lambda}-\tilde{\boldsymbol{\lambda}}$,
the error of approximation is bounded as
\begin{equation}\label{eq:ErrBound}
\|\Delta\lambda\|_{\infty} \leq \frac{\|\mathbf{D}^{-1}\|_{\infty}
\|\Delta\mathbf{A}\|_{\infty}}{1-\|\mathbf{D}^{-1}\|_{\infty}
\|\Delta\mathbf{A}\|_{\infty}}
\|\mathbf{D}^{-1}\mathbf{y}\|_{\infty}
\end{equation}
when $\|\mathbf{D}^{-1}\|_{\infty}
\|\Delta\mathbf{A}\|_{\infty}<1$. \vspace{5pt}

\noindent \textbf{Proof:} \hspace{5pt} By
Eq.(\ref{eq:eqRegularizedLinearSystemMatrix}), we have
\[
(\mathbf{D} + \Delta\mathbf{A})(\tilde{\boldsymbol{\lambda}} +
\Delta\boldsymbol{\lambda}) = \mathbf{y}.
\]
With the quasi-solution that $\tilde{\boldsymbol{\lambda}} =
\mathbf{D}^{-1} \mathbf{y}$, this equation can be converted to
\[
\Delta\boldsymbol{\lambda} = \mathbf{D}^{-1}[- (\Delta\mathbf{A})
\tilde{\boldsymbol{\lambda}} - (\Delta\mathbf{A})
(\Delta\boldsymbol{\lambda})].
\]
Then, we apply a maximum norm and get
\[
\|\Delta\boldsymbol{\lambda}\|_{\infty} \leq
\|\mathbf{D}^{-1}\|_{\infty}(\| \Delta\mathbf{A}\|_{\infty}
\|\tilde{\boldsymbol{\lambda}}\|_{\infty} +
\|\Delta\mathbf{A}\|_{\infty}
\|\Delta\boldsymbol{\lambda}\|_{\infty}),
\]
which is also
\[
(1-\|\mathbf{D}^{-1}\|_{\infty} \|
\Delta\mathbf{A}\|_{\infty})\|\Delta\boldsymbol{\lambda}\|_{\infty}
\leq \|\mathbf{D}^{-1}\|_{\infty} \| \Delta\mathbf{A}\|_{\infty}
\|\tilde{\boldsymbol{\lambda}}\|_{\infty}.
\]
By the given condition $\|\mathbf{D}^{-1}\|_{\infty} \|\Delta\mathbf{A}\|_{\infty}<1$, we have
\[
\|\Delta\lambda\|_{\infty} \leq \frac{\|\mathbf{D}^{-1}\|_{\infty}
\|\Delta\mathbf{A}\|_{\infty}}{1-\|\mathbf{D}^{-1}\|_{\infty}
\|\Delta\mathbf{A}\|_{\infty}}
\|\tilde{\boldsymbol{\lambda}}\|_{\infty}.
\]
Combining with Eq.(\ref{eq:quasiSolution}), the lemma has been
proved.

\hfill$\blacksquare$ \vspace{8pt} 

Assuming there are at most $m$ other centers falling in the support region for each kernel, the error-bound of our
quasi-solution can be achieved on the Wendsland's CSRBFs (i.e., Eq.(\ref{eq:CSRBF})).

\vspace{8pt} \noindent \textbf{Lemma 2} \hspace{5pt} When
Wendsland's CSRBFs are used, if their support sizes $\rho_i \in
[\rho_{\min},\rho_{\max}]$ (with $\rho_{\max}<\sqrt{20}$) and each
support region contains at most $m$ centers of other CSRBFs, the
error of $\|\Delta\lambda\|_{\infty}$ is bounded by a constant
when
\begin{equation}\label{eq:EtaBound}
\eta > \left( m(\frac{5}{4 \rho_{\min}}+\frac{35}{\rho_{\min}^2})
- 1 \right).
\end{equation}

\noindent \textbf{Proof:} \hspace{5pt} By the definition of the diagonal matrix $\mathbf{D}$ in Eq.(\ref{eq:D_matrix}),
\[
\|\mathbf{D}^{-1}\|_{\infty}=\max_{j=1,\ldots,n}
\{\frac{1}{1+\eta},\frac{\rho_j^2}{20+\eta \rho_j^2} \}.
\]
The upper bound of $\|\mathbf{D}^{-1} \mathbf{y}\|_{\infty}$ can
also be obtained from Eq.(\ref{eq:quasiSolution}) as
\begin{align*}
\|\mathbf{D}^{-1} \mathbf{y}\|_{\infty}&=\|
\tilde{\boldsymbol{\lambda}} \|_{\infty} \\
&=\max_{j=1,\ldots,n} \{ 0, \frac{ \rho_j^2
\mathbf{n}_j^x}{20+\eta \rho_j^2} , \frac{ \rho_j^2
\mathbf{n}_j^y}{20+\eta \rho_j^2}, \frac{ \rho_j^2
\mathbf{n}_j^z}{20+\eta \rho_j^2} \}.
\end{align*}
Here, superscripts denote the $x$-, $y$- and $z$-components of a
vector in $\Re^3$. When $\rho_i > \rho_j > 0$ and $\eta \geq 0$,
\[
\frac{\rho_i^2}{20+\eta \rho_i^2}=\frac{1}{20/\rho_i^2+\eta}
> \frac{\rho_j^2}{20+\eta \rho_j^2}=\frac{1}{20/\rho_j^2+\eta}.
\]
As a result
\begin{align*}
\|\mathbf{D}^{-1}\|_{\infty} & \leq \max
(\frac{1}{1+\eta},\frac{\rho_{\max}^2}{20+\eta \rho_{\max}^2} ) \\
\|\mathbf{D}^{-1} \mathbf{y}\|_{\infty}& \leq
\frac{\rho_{\max}^2}{20+\eta \rho_{\max}^2}.
\end{align*}
When $\rho_j \leq \rho_{\max}<\sqrt{20}$, we can further obtain
\[
\|\mathbf{D}^{-1}\|_{\infty}=\frac{1}{1+\eta}.
\]
Now we derive the upper bound of $\|\Delta\mathbf{A}\|_{\infty}$.
From Eqs.(\ref{eq:coefficientMatrixBlocks}) and
(\ref{eq:D_matrix}), using $\varphi_{i,j}$ to denote
$\varphi_i(\mathbf{p}_j)=\varphi(\mathbf{p}_j-\mathbf{p}_i)$, we
can also have
\begin{align*}
&\|\Delta\mathbf{A}\|_{\infty} = \max_{j=1,...,n}\{ \\
&\sum_{i}^{m}(|\varphi_{i,j}|+|\frac{\partial \varphi_{i,j}}{\partial x}|+|\frac{\partial \varphi_{i,j}}{\partial
y}|+|\frac{\partial \varphi_{i,j}}{\partial z}|),
\\
&\sum_{i}^{m}(|\frac{\partial \varphi_{i,j}}{\partial x}|+|\frac{\partial^2 \varphi_{i,j}}{\partial
x^2}|+|\frac{\partial^2 \varphi_{i,j}}{\partial x\partial y}|+|\frac{\partial^2 \varphi_{i,j}}{\partial x\partial z}|),
\\
&\sum_{i}^{m}(|\frac{\partial \varphi_{i,j}}{\partial y}|+|\frac{\partial^2 \varphi_{i,j}}{\partial x\partial
y}|+|\frac{\partial^2 \varphi_{i,j}}{\partial y^2}|+|\frac{\partial^2 \varphi_{i,j}}{\partial y\partial z}|),
\\
&\sum_{i}^{m}(|\frac{\partial \varphi_{i,j}}{\partial z}|+|\frac{\partial^2 \varphi_{i,j}}{\partial x\partial
z}|+|\frac{\partial^2 \varphi_{i,j}}{\partial y\partial z}|+|\frac{\partial^2 \varphi_{i,j}}{\partial z^2}|)\}.
\end{align*}
By the derivatives listed in Table \ref{tabCSRBFDerivatives} and their corresponding upper bounds listed in Table
\ref{tabCSRBFDerivativesBnd}, we can have
\begin{align*}
\|\Delta\mathbf{A}\|_{\infty} & \leq \max_j \{ m(1+\frac{15}{4
\rho_j}), m(\frac{5}{4 \rho_j}+\frac{35}{\rho_j^2}) \} \\
& \leq \max \{m(1+\frac{15}{4 \rho_{\min}}), m(\frac{5}{4 \rho_{\min}}+\frac{35}{\rho_{\min}^2}) \}.
\end{align*}
When $\rho_{\min} \leq \rho_{\max} < \sqrt{20}$, it can easily be
further simplified to
\[
\|\Delta\mathbf{A}\|_{\infty} \leq m(\frac{5}{4 \rho_{\min}}+\frac{35}{\rho_{\min}^2}) \equiv \bar{A}.
\]
Summarizing all the analysis together, we have
\begin{equation}\label{eq:ErrBoundConst}
\|\Delta\lambda\|_{\infty} \leq \frac{ \bar{A}
\rho_{\max}^2}{(1+\eta - \bar{A})(20 + \eta \rho_{\max}^2)}
\end{equation}
when $\|\mathbf{D}^{-1}\|_{\infty} \|\Delta\mathbf{A}\|_{\infty} \leq \bar{A}  / (1+\eta) < 1$. To hold this, it should
have $1+\eta> \bar{A}$ -- that is Eq.(\ref{eq:EtaBound}). The lemma has been proved.

\hfill$\blacksquare$ \vspace{8pt} 

\begin{table}
\centering \caption{Derivatives of Wendsland's
CSRBF~\cite{Wendland:1995}}\label{tabCSRBFDerivatives} {
\renewcommand{\arraystretch}{2.5}
\begin{tabular}{|c|l|}
  \hline
  $\displaystyle \varphi_i(\mathbf{x})$ & $\displaystyle (1-\frac{r}{\rho_i})^4(4\frac{r}{\rho_i}+1)$ \\
  \hline
  $\displaystyle \frac{\partial \varphi_i(\mathbf{x})}{\partial x}$ & $\displaystyle -\frac{20}{\rho_i^2}(1-\frac{r}{\rho_i})^3(x-x_i)$ \\
  $\displaystyle \frac{\partial \varphi_i(\mathbf{x})}{\partial y}$ & $\displaystyle -\frac{20}{\rho_i^2}(1-\frac{r}{\rho_i})^3(y-y_i)$ \\
  $\displaystyle \frac{\partial \varphi_i(\mathbf{x})}{\partial z}$ & $\displaystyle -\frac{20}{\rho_i^2}(1-\frac{r}{\rho_i})^3(z-z_i)$ \\
  $\displaystyle \frac{\partial^2 \varphi_i(\mathbf{x})}{\partial x^2}$ & $\displaystyle -\frac{20}{\rho_i^2}(1-\frac{r}{\rho_i})^3+\frac{60}{\rho_i^3}(1-\frac{r}{\rho_i})^2\frac{(x-x_i)^2}{r}$ \\
  $\displaystyle \frac{\partial^2 \varphi_i(\mathbf{x})}{\partial y^2}$ & $\displaystyle -\frac{20}{\rho_i^2}(1-\frac{r}{\rho_i})^3+\frac{60}{\rho_i^3}(1-\frac{r}{\rho_i})^2\frac{(y-y_i)^2}{r}$ \\
  $\displaystyle \frac{\partial^2 \varphi_i(\mathbf{x})}{\partial z^2}$ & $\displaystyle -\frac{20}{\rho_i^2}(1-\frac{r}{\rho_i})^3+\frac{60}{\rho_i^3}(1-\frac{r}{\rho_i})^2\frac{(z-z_i)^2}{r}$ \\
  $\displaystyle \frac{\partial^2 \varphi_i(\mathbf{x})}{\partial x \partial y}$ & $\displaystyle \frac{60}{\rho_i^3}(1-\frac{r}{\rho_i})^2\frac{(x-x_i)(y-y_i)}{r}$ \\
  $\displaystyle \frac{\partial^2 \varphi_i(\mathbf{x})}{\partial x \partial z}$ & $\displaystyle \frac{60}{\rho_i^3}(1-\frac{r}{\rho_i})^2\frac{(x-x_i)(z-z_i)}{r}$ \\
  $\displaystyle \frac{\partial^2 \varphi_i(\mathbf{x})}{\partial y \partial z}$ & $\displaystyle \frac{60}{\rho_i^3}(1-\frac{r}{\rho_i})^2\frac{(y-y_i)(z-z_i)}{r}$ \\
  \hline
\end{tabular}
}
\begin{flushleft}
$^{\dagger}$Here, $\mathbf{x}=(x,y,z)$,
$r=\sqrt{(x-x_i)^2+(y-y_i)^2+(z-z_i)^2}$ and $\rho_i$ is the
support size of the radial basis function $\varphi_i(\mathbf{x})$.
\end{flushleft}
\end{table}

\begin{table}
\centering  \caption{Error Bounds of
Derivatives}\label{tabCSRBFDerivativesBnd}{
\renewcommand{\arraystretch}{2.5}
\begin{tabular}{|c|l|}
  \hline
  $\displaystyle |\varphi_i(\mathbf{x})|$ & $\displaystyle \leq 1$ \\
  \hline
  $\displaystyle |\frac{\partial \varphi_i(\mathbf{x})}{\partial x}|, \; |\frac{\partial \varphi_i(\mathbf{x})}{\partial x}|, \; |\frac{\partial \varphi_i(\mathbf{x})}{\partial x}|$
    & $\displaystyle \leq \frac{20}{\rho_i}(1-\frac{r}{\rho_i})^3 \frac{r}{\rho_i}$  \\
    & $\displaystyle \leq \frac{5}{4\rho_i}$ \;\; (with $r=\frac{\rho_i}{2}$)\\
  \hline
  $\displaystyle |\frac{\partial^2 \varphi_i(\mathbf{x})}{\partial x^2}|, \; |\frac{\partial^2 \varphi_i(\mathbf{x})}{\partial y^2}|, \; |\frac{\partial^2 \varphi_i(\mathbf{x})}{\partial z^2}|,$
    & $\displaystyle \leq \frac{20}{\rho_i^2}(1-\frac{r}{\rho_i})^2(1+2\frac{r}{\rho_i}) $\\
    & $\displaystyle \leq \frac{20}{\rho_i^2}$ \;\; (with $r=0$)\\
  \hline
  $\displaystyle |\frac{\partial^2 \varphi_i(\mathbf{x})}{\partial x \partial y}|, \; |\frac{\partial^2 \varphi_i(\mathbf{x})}{\partial x \partial z}|, \; |\frac{\partial^2 \varphi_i(\mathbf{x})}{\partial x \partial y}|,$
    & $\displaystyle \leq \frac{60}{\rho_i^2}(1-\frac{r}{\rho_i})^2 \frac{r}{\rho_i}$ \\
    & $\displaystyle \leq \frac{15}{2\rho_i^2}$ \;\; (with $r=\frac{\rho_i}{2}$) \\
  \hline
\end{tabular}
}
\begin{flushleft}
$^{\ddagger}$The analysis is based on $|x-x_i|\leq r$,
$|y-y_i|\leq r$ and $|z-z_i|\leq r$, and the bound is derived by
using the inequality of arithmetic and geometric means.
\end{flushleft}
\end{table}

\noindent \textit{Remark.} \hspace{5pt} The requirements of,
\begin{enumerate}
\item all the CSRBFs have their support sizes within the interval $[\rho_{\min},\rho_{\max}]$, and

\item there are at most $m$ centers falling in the support of any others CSRBF,
\end{enumerate}
can be achieved by a carefully designed parameter tuning algorithm (see Section \ref{subsecSupportSize}). After
determining the support sizes, the value of $\eta$ can be chosen to hold Eq.(\ref{eq:EtaBound}). By scaling all models
into a bounding box of $[-1,1]^3 \in \Re^3$, the algorithm in Section \ref{subsecSupportSize} can also enforce
$\rho_{\max}<\sqrt{20}$. Numerical errors generated in our experimental tests are discussed in Section
\ref{subsecErrVerification}, which further verify the error-bound of our method.

%
%
\section{Reconstruction Algorithm}\label{secReconstruction}
Given points and their normals in the input sets $\mathcal{P}$ and $\mathcal{N}$, the implicit function
$\tilde{f}(\mathbf{x})$ defined in Eq.(\ref{eq:appFuncClosedForm}) can be evaluated in the supported regions as an
approximation of the HRBF reconstruction. Specifically, with the help of tessellation techniques, zero isosurface of
the implicit function $\tilde{f}(\mathbf{x})$ can be converted into a mesh surface. A scheme is also developed to
determine the support sizes of HRBFs and the coefficient $\eta$ according to \textit{Lemma~2} to guarantee the
existence of error-bound.

Our reconstruction algorithm consists of a scheme to tune parameters according to the analysis for error-bound and an
efficient method akin to DC to extract the zero isosurface as a polygonal mesh in the supported regions of
$\tilde{f}(\mathbf{x})$. When processing highly non-uniform point sets, an optional step of center selection is needed
to reduce the artifacts caused by the non-uniformity. After that, the input sets $\mathcal{P}$ and $\mathcal{N}$ are
reduced into a set $\mathcal{C}=\{ \mathbf{c}_1, \mathbf{n}_1, \cdots, \mathbf{c}_l, \mathbf{n}_l \}$ with less Hermite
points (i.e., $l<n$).

\subsection{Parameters tuning}\label{subsecSupportSize}
The support sizes of HRBF implicits, $\rho_j$s, and the coefficient of regularization, $\eta$, should satisfy the
condition in Eq.(\ref{eq:EtaBound}) for the existence of error-bound. Specifically, the following factors must be
considered:
\begin{itemize}
\item $\rho_{\max}=\max\{\rho_j\} < \sqrt{20}$ gives the upper bounds of $\rho_j$.

\item According to Eq.(\ref{eq:EtaBound}), with larger $\rho_{\min}$, users will have more flexibility to choose the
value of $\eta$ (i.e., has smaller value for the right of the inequality in Eq.(\ref{eq:EtaBound})).

\item On the other aspect, if increasing the support size $\rho_j$ of a RBF centered at $\mathbf{p}_j$, more other
centers will be covered by the supporting region. This increases the value of $m$ in \textit{Lemma 2} and thus the
right-hand side of the inequality in Eq.(\ref{eq:EtaBound}).
\end{itemize}
Based on these reasons, we develop a scheme below to determine the values of $\rho_j$s and $\eta$.

Each support must cover enough number of data points to generate a span of the local shape; meanwhile, it cannot be too
large. For this purpose, we first construct an octree to split the input points into different nodes by keeping similar
number of points in each leaf-nodes. $\bar{d}$ is then set as $3/4$ of the average diagonal length of the leaf-nodes.
The support sizes are temporally set as $s \bar{d}$ with $s$ being an amplifier -- $s=1.0$ is employed for clean data
sets. We then count the number of points covered by the supporting region of each CSRBF, $\phi_{\rho_j}(\|\mathbf{x} -
\mathbf{p}_j\|)$. The value of $m$ is selected as the maximal number of data points covered by each of these temporary
supports. After determining the value of $m$, $\rho_j$ of each CSRBF is enlarged by an incremental procedure until the
support contains more than $m$ points. By the fixed support sizes, the value of $\eta$ can be chosen by
Eq.(\ref{eq:EtaBound}). In all our tests, we assign it with a value slightly larger (e.g., $10^{-5}$) than the
right-hand side of Eq.(\ref{eq:EtaBound}).

\begin{figure*}
\centering
\includegraphics[width=\linewidth]{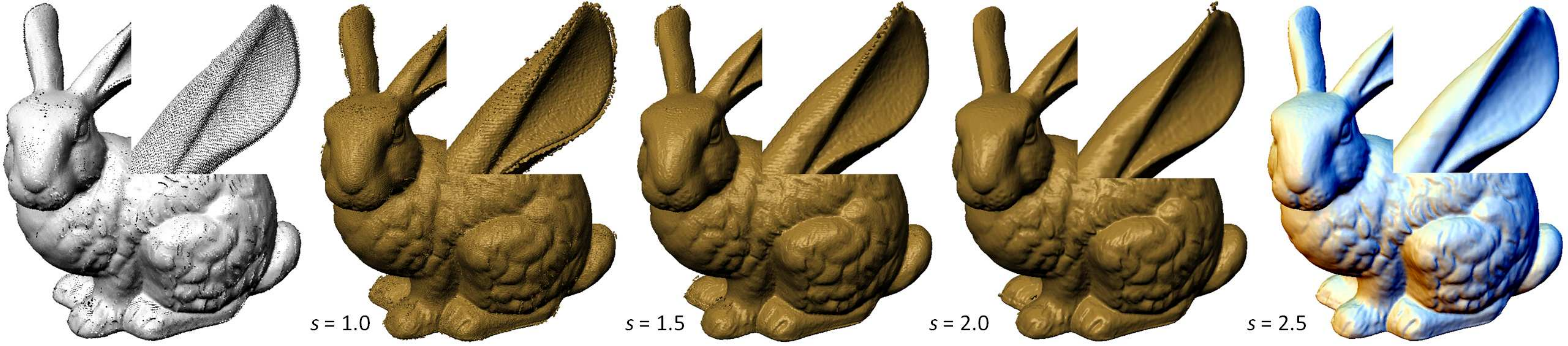}
\caption{Results with different regularization can be obtained by using different amplifiers on a noisy point set:
(from left to right) input noisy points and our reconstruction results. In these examples, $\eta$ is chosen as a value
slightly greater than the right of Eq.(\ref{eq:EtaBound}).}\label{fig:RegularizationList}
\end{figure*}

The values of $\rho_j$s determined by the above method work well on clean data but may fail in highly noisy input.
Further tuning is needed. First of all, we enlarge the temporal support sizes by using $s>1.0$ to enhance the
effectiveness of denoising. This results in a larger $m$. More geometric details can be preserved with a smaller
support size while a larger support size leads to a smoother reconstruction. Moreover, among all the $\rho_j$s obtained
by the aforementioned method, the minimal value of $\rho_j$s is selected as the final support size for all CSRBF
kernels -- i.e., uniform support size is adopted for highly noisy input. Examples using different amplifiers can be
found in Fig.\ref{fig:RegularizationList}.

\begin{figure*}
\includegraphics[width=\linewidth]{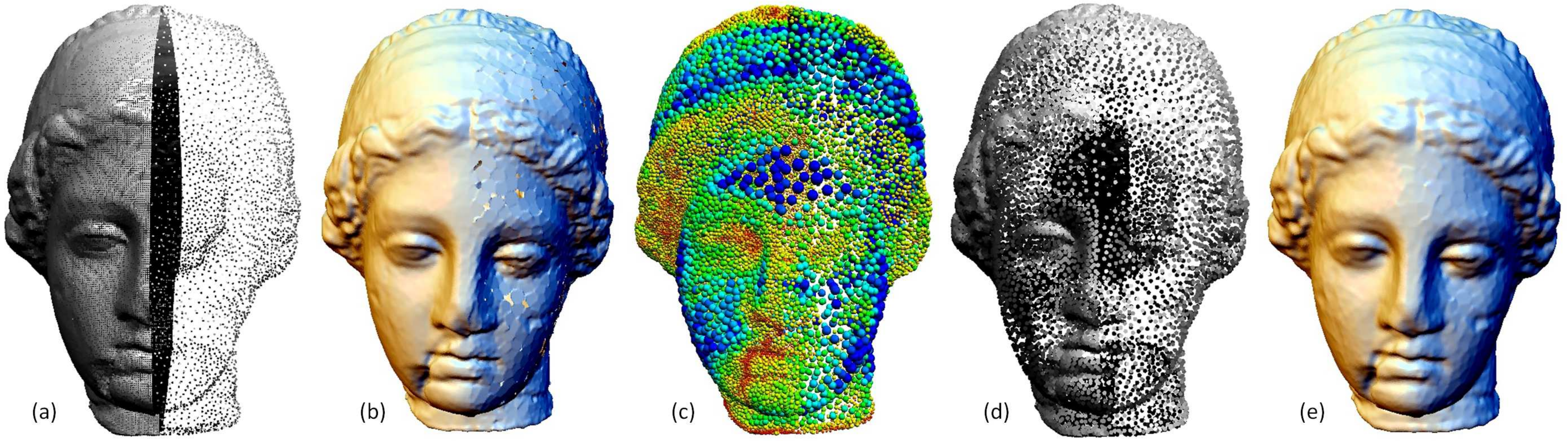}
\caption{Adaptive HRBF implicits are generated by our method with the help of center selection: (a) the input set with
$100,371$ points in high non-uniformity, (b) the reconstruction using all points as centers of HRBF implicits will
easily lead to holes in the sparse regions, (c) the spherical cover -- the spheres are displayed in radii as $1/4$ of
the real ones, (d) the selected $13,446$ centers of RBFs, and (e) reconstruction from the selected centers -- no hole
will be generated as the densities of centers in the left and the right are similar to each other. The support sizes
are determined by the method in Section \ref{subsecSupportSize} with (b) $s=2.1$ and (e) $s=1.0$, both of which lead to
$m=120$.} \vspace{-10pt}\label{fig:CenterSelectionToReduceArtifact}
\end{figure*}

\subsection{Efficient isosurface extraction}\label{subsecIsosurfaceExtraction}
Our surface reconstruction method only evaluates the function values of the implicit function $\tilde{f}(\mathbf{x})$
during the isosurface extraction step. An variation of DC algorithm~\cite{Ju:2002} is developed to extract zero
isosurfaces from the regions spanned by the RBFs in Eq.(\ref{eq:appFuncClosedForm}). In our method, the reconstruction
with compactly supported implicit functions leads to open mesh surfaces and leaves holes in regions not covered by the
supports of RBFs. This is very useful for reconstructing the scenes that have not been completely captured (e.g., the
scene shown in Fig.\ref{fig:IndoorReconstructionExample}). Moreover, as $\tilde{f}(\mathbf{x})$ is defined in a
closed-form, the function evaluation (therefore also the isosurface extraction) is highly scalable and can be
efficiently performed by local operations.

Voxels with a fixed width $w$ are constructed and those intersecting the isosurface $\tilde{f}(\mathbf{x})=0$ are first
searched in the supported regions around the centers $\{\mathbf{p}_j\}$. The voxels are constructed only when 1) all of
its eight corners have function values \textit{defined} and 2) the function values have \textit{different signs}. For
each edge $e$ of a voxel, the intersection between $e$ and the isosurface can be determined by a bi-sectional search
when two endpoints of $e$ have different signs in $\tilde{f}(\cdot)$. The normal of this intersection point is assigned
as $\nabla \tilde{f}$. For each voxel, a vertex is constructed at the position $\mathbf{v}$ that minimize the
quadratic-function $\sum_j ((\mathbf{v}-\mathbf{q}_j)\cdot \mathbf{n}_{q_j})^2$, where $\mathbf{q}_j$ and
$\mathbf{n}_{q_j}$ are the intersection points on edges and the normal vectors at the intersections. For each edge $e$
with intersection, a quadrangle is constructed by linking the vertices in the four voxels around $e$. As a result, the
final surface mesh can be obtained.

In our current implementation, the mesh surface is extracted from voxels with a fixed width. More sophisticated
algorithm can be developed to construct an octree based hierarchy to extract triangles adaptively on a partial region
of implicit surfaces (e.g., \cite{Wang:2011}). All the operations in our isosurface evaluation and extraction steps are
local. It is easy to be implemented in parallel on many-cores or in a distributed environment. By the algorithm's
locality, a progressive reconstruction can be easily developed by only updating a local region when new points (as
centers of CSRBFs) are added.

\subsection{Center selection}\label{subsecCenterSelection}
This is an optimal step to be applied when high non-uniformity is observed on the input points. For such a point set,
the direct reconstruction by using all points as centers of CSRBFs results in a reconstruction with holes in the sparse
regions (see Fig.\ref{fig:CenterSelectionToReduceArtifact}(b) for an example). Here, we adaptively select samples from
$\mathcal{P}$ and $\mathcal{N}$ to form a subset $\mathcal{C}$. The Hermite points in $\mathcal{C}$ will be used as
centers of HRBF in the above method to obtain a better surface reconstruction. Each center, $\mathbf{c}_k$, is also
associated with a radius, $r_k$, which therefore forms a local spherical cover of the given points.

\vspace{8pt} \noindent \textbf{Definition 2}~~The \textit{degree of coverage} (DoC) at a point $\mathbf{x} \in \Re^3$
is defined as a function
\begin{equation}\label{eq:eqDegreeOfCoverage}
g(\mathbf{x})=\sum_{k=1}^l { \phi_{r_k}(\| \mathbf{x} - \mathbf{c}_k \|) }
\end{equation}
according to a set of down-sampled Hermite points, $\mathcal{C}=\{ \mathbf{c}_1, \mathbf{n}_1, \cdots, \mathbf{c}_l,
\mathbf{n}_l \}$. \vspace{8pt}

\noindent We wish to generate a minimal spherical cover by controlling DoC in an iterative procedure.

The basic idea of center selection is to form spherical covers by letting DoC at every point in $\mathcal{P}$ not less
than a criterion $g_{\min}$ (i.e., $\forall \mathbf{p}_j \in \mathcal{P}, \; g(\mathbf{p}_j) \geq g_{\min}$). To this
end, the following steps are iteratively run until the criterion is satisfied at all points.
\begin{enumerate}
\item In the initial step, $\mathcal{C}=\emptyset$ and $g_j=0$ is assigned to all points $\mathbf{p}_j \in
\mathcal{P}$.

\item Randomly selecting $\varpi$ points with their DoCs less than $g_{\min}$. Among these $\varpi$ points, the point
with the smallest $g(\cdot)$ is chosen as a center $\mathbf{c}_k$ to add into $\mathcal{C}$ together with its normal
vector.

\item The radius $r_k$ of sphere centered at $\mathbf{c}_k$ is then determined by a quadric-error function
\begin{equation*}
q(\mathbf{c}_k,r_k)=\frac{ \sum_j {\delta_j \phi_{r_k} (\| \mathbf{p}_j - \mathbf{c}_k \|) (\mathbf{n}_j \cdot
(\mathbf{c}_k - \mathbf{p}_j))^2 } } { \sum_j {\delta_j \phi_{r_k} (\| \mathbf{p}_j - \mathbf{c}_k \|) } },
\end{equation*}
which evaluates how curved the surface inside the sphere is -- the shape is represented by sample points in
$\mathcal{P}$. In other words, for a highly curved region, a sphere with smaller $r_k$ should be used to reduce the
error. Here, $\delta_j$ is the average of the squared distances between a point $\mathbf{p}_j$ and its 15 nearest
neighboring points. The value of $\delta_j$ indicates a weight of point density. The bi-sectional search is taken to
obtain a maximal $r_k$ that still satisfies
\begin{center}
$q(\mathbf{c}_k,r_k) \leq q_{err} \bar{L}$
\end{center}
with $\bar{L}$ being the diagonal length of the input points'
bounding box.

\item Updating DoC at all points $\mathbf{p}_j$ within the range $\|\mathbf{p}_j - \mathbf{c}_k \| < r_k$ while
$g(\mathbf{p}_j)<g_{\min}$. DoC of $\mathbf{c}_k$ is assigned as $g_{\min}$ to avoid being selected as candidates of
centers once again.

\item Go back to step 2) until DoC at all points are not less than $g_{\min}$.
\end{enumerate}
Note that, this iterative procedure is a variant of our prior work in \cite{Liu:2010} with certain modification to fit
the formulation of CSRBF. The efficiency of computation has also been improved. An example result of our minimal
spherical covering is shown in Fig.\ref{fig:CenterSelectionToReduceArtifact}(c), where the selected centers of RBFs are
displayed as spheres. Colors are used to represent the sizes of spheres with red for the smallest and blue for the
biggest ones. Samples adaptive to the geometric details have been illustrated as
Fig.\ref{fig:CenterSelectionToReduceArtifact}(d). In all examples of this paper, $g_{\min}=1.5$, $q_{err}=5 \times
10^{-4}$ and $\varpi=15$ work well.

\begin{figure*}
\includegraphics[width=\textwidth]{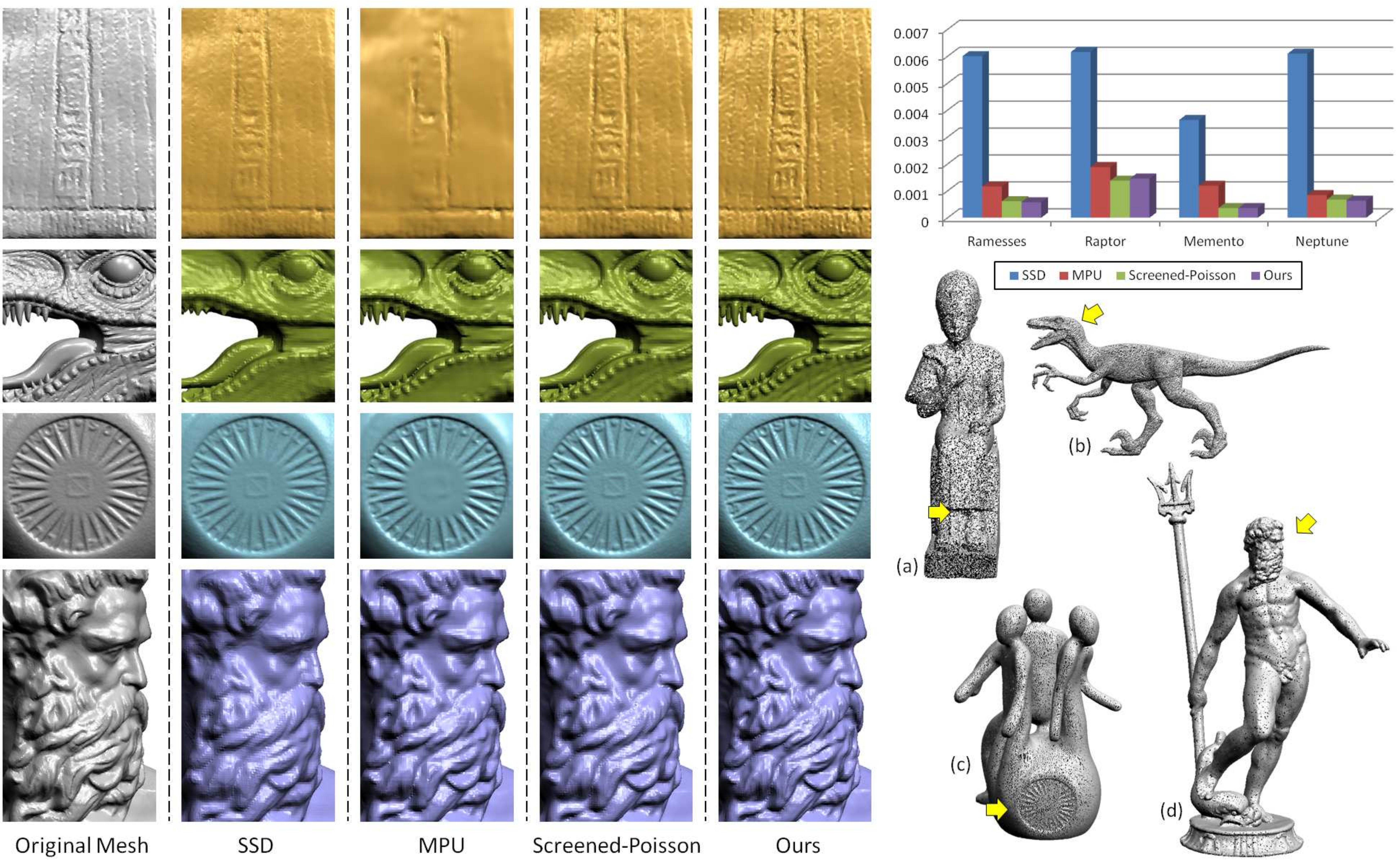}
\caption{Experimental tests on clean point cloud that is uniformly sampled from fours mesh models -- (a) Ramesses
($0.580M$ pts.), (b) Raptor ($1.00M$ pts.), (c) Momento ($2.52M$ pts.) and (d) Neptune ($4.98M$ pts.). For
illustration, only $1/10$ points are displayed for the points of the first three models, and $1/20$ points for the
Neptune model. The reconstructions by different methods including SSD, MPU, Screened-Poisson and ours are shown and
compared in the left. A bar chat is also given to report the average shape approximation errors on different
reconstructions by the Metro tool~\cite{Cignoni:1998}. We use $s\bar{d}=0.003$ as the support size and $w=0.001$ for
the grid width of polygonization in all examples here. To conduct a fair comparison, similar number of triangles are
generated through the polygonization for different approaches.} \vspace{-10pt}\label{fig:CleanResults}
\end{figure*}

With the centers selected above, a reduced implicit function can be obtained by Eq.(\ref{eq:appFuncClosedForm}) but
with fewer centers of CSRBFs. After applying the center selection step, the density of centers at each region becomes
compatible to its neighboring regions -- i.e., no sharp change. As a result, a better reconstruction can be obtained
(see the result shown in Fig.\ref{fig:CenterSelectionToReduceArtifact}(e)).

\section{Results and Discussion}\label{secResult}
An surface reconstruction algorithm based on the closed-form formulation of HRBF implicits has been implemented with
Microsoft Visual C++ and OpenGL. We evaluate our methods on a PC with two Intel Core i7-2600K CPUs at 3.4GHz plus 16GB
RAM. Our approach has been applied to various data with up to fourteen millions of points (the surface can be
reconstructed in $78.9$ sec.). Our results are compared with a variety of approaches -- see the examples and
discussions presented below. All the models are re-scaled into a bounding-box of $[-1,1]^3\in \Re^3$.

\subsection{Comparisons}
Firstly, we test the performance of our approach on sets of clean points, which are uniformly sampled from polygonal
meshes. Four models, Ramesses, Raptor, Momento and Neptune, are sampled into sets with $0.58M \sim 4.98M$ points. Our
results are compared with three prior methods, including the \textit{Multiple Partition of Unity} (MPU) reconstruction
~\cite{Ohtake:2003}, the \textit{Smooth Signed Distance} (SSD) reconstruction~\cite{Calakli:2011} and the
\textit{Screened-Poisson} reconstruction~\cite{Kazhdan:2013}. Comparisons are shown in Fig.\ref{fig:CleanResults}. We
employ the 10-th depth of octree in the SSD and Screen-Poisson methods to generate results in
Fig.\ref{fig:CleanResults}. For MPU and our method, we adjust the resolutions of polygonization methods to extract
meshes with similar numbers of triangles as SSD and Screened-Poisson. The parameter Max\_Error of MPU is set as $0.001$
times of the model's size. Default values are used for other parameters. From observation, it can be found that
geometric details on the original mesh can be well preserved by our method while being smoothed out in some prior
methods. Publicly available software, Metro tool~\cite{Cignoni:1998}, is employed to compute the average shape
approximation error between the reconstructed surface and the original mesh. A bar chat of errors is given in the
upper-right of Fig.\ref{fig:CleanResults}. Our method can always generate more accurate results than SSD and MPU.
Meanwhile, our results have similar accuracy comparing to Screened-Poisson.

Table~\ref{tab:StatisticsOnCleanData} gives the computational statistics of tests on these models. Due to the
close-form formulation, our method does not need any global operation such as solving a large linear system. Therefore,
its computational time is only spent on constructing an octree to computing the support size and the step of function
value evaluation in iso-surface extraction. Both SSD and Screened-Poisson need to solve linear systems globally. In
Poisson reconstruction, the multi-grid solver performs a constant number of conjugate-gradient iterations at each
level, which gives linear complexity w.r.t to the number of nodes in the octree. The SSD reconstruction uses
conjugate-gradients to determine all the coefficients simultaneously, which has a complexity of $O(n^{1.5})$. This
leads to a significantly slower performance on models with large number of points (see
Table~\ref{tab:StatisticsOnCleanData}). In MPU reconstruction, only local fitting is taken at leaf-nodes of an octree.
These surfaces are blended together to form the resultant surface, which is fast but still slower than ours. Moreover,
our method generates results with smaller shape approximation error than MPU (see Fig.\ref{fig:CleanResults}). In
summary, our method is the fastest method and can generate similar results as the best of other three in terms of
quality.

\begin{table}
\small \caption{Runtime performance of different reconstruction approaches on clean point
sets$^{\dagger}$}\label{tab:StatisticsOnCleanData} 
\begin{tabular}{|l|c|r|r|r|r|}
  \hline
  & & \multicolumn{4}{c|}{\textbf{Time in Seconds}$^{*}$} \\
  \cline{3-6}
  \textbf{Model} & \textbf{Pts.}          & \textbf{SSD} & \textbf{MPU} & \textbf{Poisson} & \textbf{Ours} \\
  \hline
  \hline
  Ramesses & 0.58M     & 14,314 & 61.2 & 40.8 & 8.3 \\
  Raptor & 1.00M        & 1,799 & 47.2 & 31.6 & 6.8 \\
  Memento & 2.52M       & 24,195 & 138.8 & 92.6 & 20.4 \\
  Neptune & 4.98M       & 6,772 & 139.4 & 114.0 & 18.9 \\
  \hline
\end{tabular}
\begin{flushleft}
$^{*}$Note that, the time reported here includes both the surface reconstruction and the mesh extraction. \\
$^{\dagger}$To have a fair comparison, similar number of triangles are generated for different approaches.
\end{flushleft}\vspace{-10pt}
\end{table}

\subsection{Raw data}
\begin{figure*}
\includegraphics[width=\textwidth]{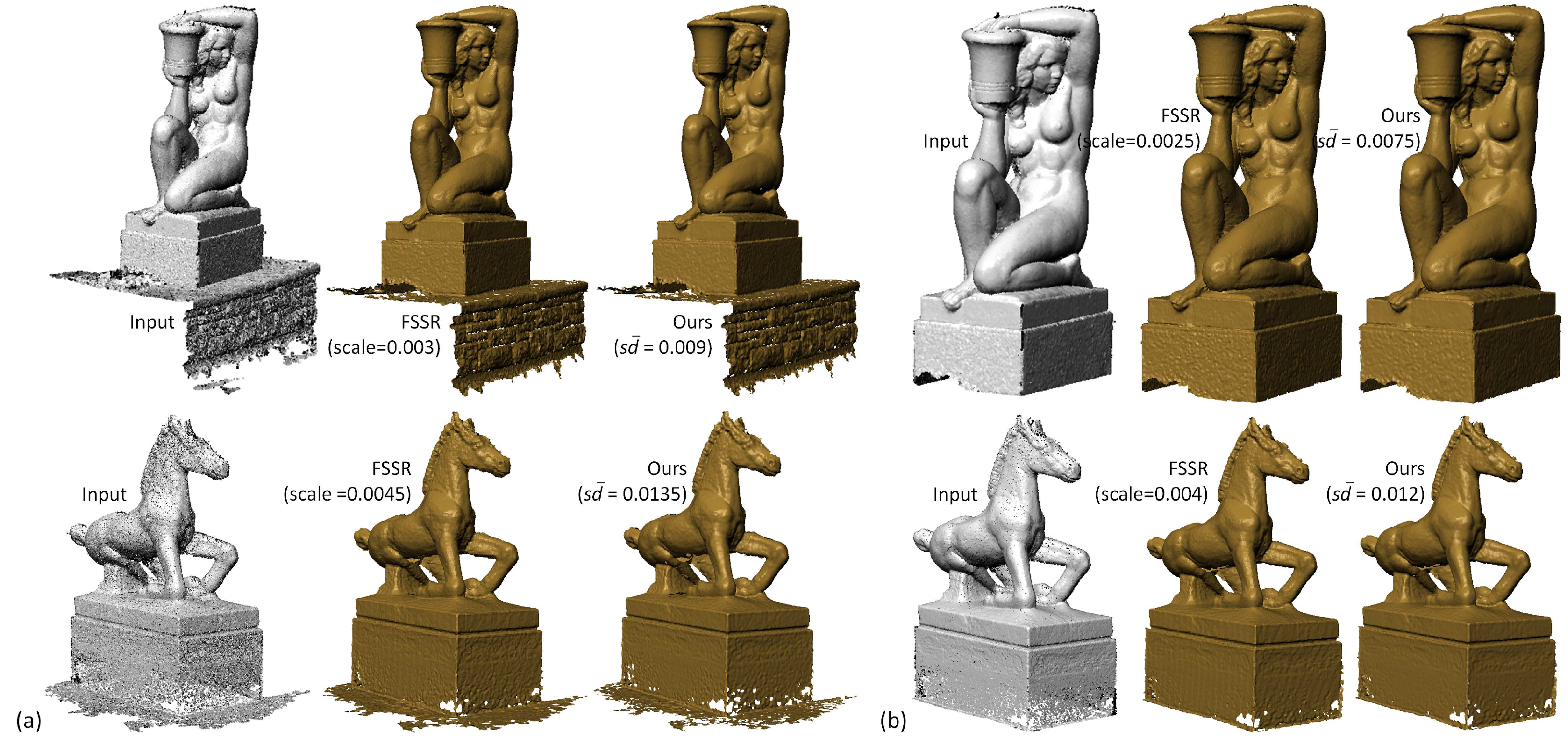}\\ \small \centering 
\begin{tabular}{|c|c|c|c|c|c|c|c|}
  \hline
  & & \multicolumn{3}{c|}{\textbf{FSSR}} & \multicolumn{3}{c|}{\textbf{Ours}} \\
  \cline{3-8}
             & \textbf{Num. of  Points} & \textbf{Num. of} & \multicolumn{2}{c|}{\textbf{Time in Seconds}} & \textbf{Num. of} & \multicolumn{2}{c|}{\textbf{Time in Seconds}} \\
  \cline{4-5} \cline{7-8}
  \textbf{Model} & \textbf{After Consolidation} & \textbf{Triangles} & One-core & 8-cores & \textbf{Triangles} & One-core & 8-cores \\
  \hline
  Aquarius  &  833.4k & 624,916 & 826.9 &  172.3  & 624,363  & 82.5 &   19.5 \\
  Horse &  239.8k & 241,614 & 263.7 &  56.4 &   623,772  & 37.3  &   9.8 \\
  \hline
\end{tabular}
\caption{Examples of surface reconstruction on incomplete set of points: (a) reconstruction from raw data and (b)
reconstruction from data sets processed by the consolidation method~\cite{Wang:2013}. Our results are comparable with
that obtained by FSSR but ours is $5.76\times \sim 10.0\times$ faster.} \vspace{-10pt}\label{fig:IncompleteDataResults}
\end{figure*}

\begin{figure*}
\includegraphics[width=\linewidth]{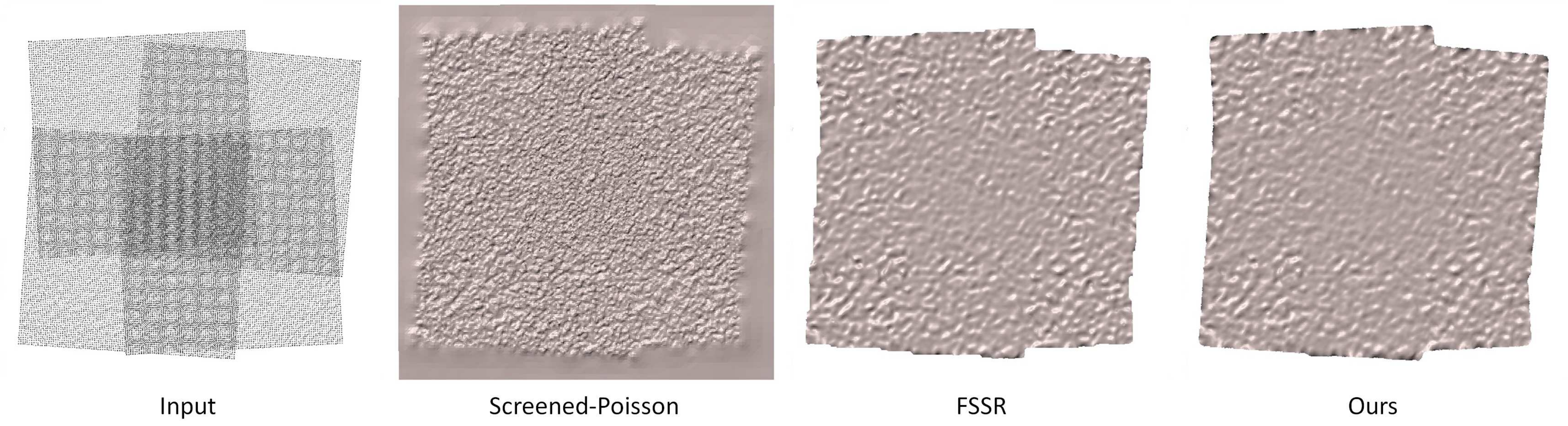}
\caption{When processing an input with significant density variation -- e.g., from four synthetic scans (most-left),
FSSR and ours can avoid generating unwanted artifacts caused by high frequency noises. The total time of our
reconstruction is 6.81~sec. ($s=3.0$ and 342k triangles are obtained on the resultant mesh), while FSSR takes 670~sec.
and results in 301k triangles (scale=0.0105). Both are tested on a CPU with eight-cores.
}\label{fig:FourSyntheticScansResult}
\end{figure*}

In practice, the real data obtained from an acquisition process are usually large and noisy point sets. Meanwhile, the
data sets are incomplete in most cases (e.g., the Aquarius and the Horse models shown in
Fig.\ref{fig:IncompleteDataResults}). The recently developed FSSR method~\cite{Fuhrmann:2014} is targeting on fast
surface reconstruction from such kind of real data. We compare our method with FSSR on two sets of real scanned data in
Fig.\ref{fig:IncompleteDataResults}. Fuhrmann and Goesele~\cite{Fuhrmann:2014} assumed the \textit{scale} of an input
point set is known, which however is not the case here. Although it can be computed by using the average distance to
$k$-nearest neighbors, it is difficult to set a proper $k$ for reconstructing a smooth surface. In order to make an
appropriate comparison, we use $1/3$ of the average support size determined by our method as the scale used in FSSR.
This is consistent with the formulation presented in~\cite{Fuhrmann:2014}, where the support size is set as three times
of the input scale. The actual values of scale and $s \bar{d}$ are also given in Fig.\ref{fig:IncompleteDataResults}.
It can be found that similar number of triangles are generated in FSSR and our method by setting the value of scale in
this way. Note that, as both FSSR and ours do not need any global operations during the computation of surface
reconstruction, it can easily be parallelized on the PC with multi-cores -- OpenMP is used in our implementation. We
test both approaches on a PC with 8-cores. As shown in the computational statistics in
Fig.\ref{fig:IncompleteDataResults}, the program can be effectively speed up on 8-cores. The multi-core version of FSSR
is provided by the authors on their homepage.

We also study the effectiveness of our approach on the benchmark of FSSR with shape density variation caused by
superposing point sets obtained from multiple scans. As shown in Fig.\ref{fig:FourSyntheticScansResult}, the input set
from four synthetic scans is downloaded from the homepage of FSSR's authors. Our reconstruction is similar to the
result from FSSR but ours method is much faster.

\begin{figure*}
\includegraphics[width=\textwidth]{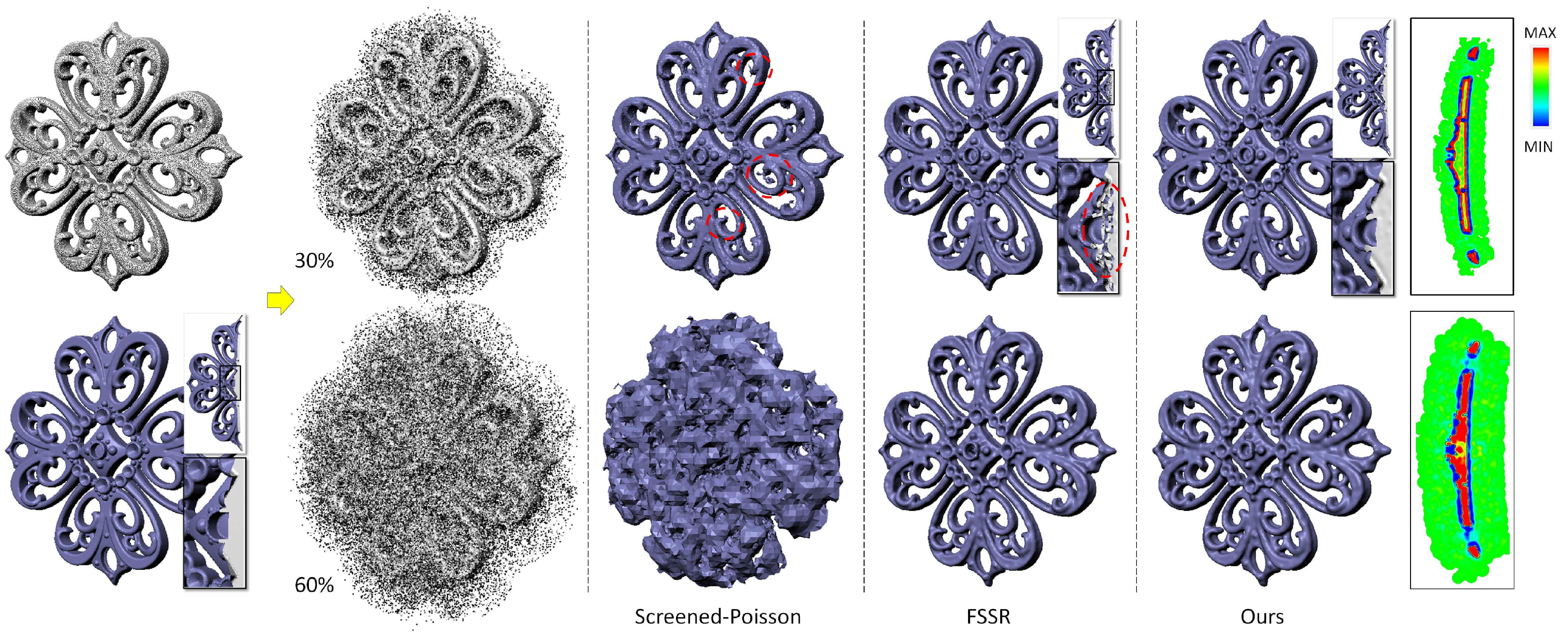}
\caption{Examples of reconstruction from sets (having 250k points) with different level of Gaussian noises. Topological
errors can be found on the results of Screen-Poisson and FSSR (see the regions circled by dashed lines in red). The
cross-sectional view of function values in our reconstruction has also been given in the right, where the regions in
white color have undefined function values.}\vspace{-10pt}\label{fig:NoisyExampleResults}
\end{figure*}

\subsection{Verification of numerical error}\label{subsecErrVerification}
Error bound of the quasi-solution $\tilde{\boldsymbol{\lambda}}$ obtained by our closed-form formulation with reference
to the exact solution $\boldsymbol{\lambda}$ of Eq.(\ref{eq:eqRegularizedLinearSystemMatrix}) has been derived in
Section \ref{subsecErrorBound}. It is also interesting to study the error between $\tilde{\boldsymbol{\lambda}}$ and
$\boldsymbol{\lambda}$. We measure $\| \tilde{\boldsymbol{\lambda}} - \boldsymbol{\lambda} \|_{\infty}$ in examples
shown above and the results are listed in Table~\ref{tab:StatisticsErr}.
\begin{table}[t]
\small \caption{Error Statistics of Quasi-Solution}\label{tab:StatisticsErr} 
\centering
\begin{tabular}{|l|c|c|c|}
  \hline
  \textbf{Model} & \textbf{Figure} & $\eta$ & $\| \tilde{\boldsymbol{\lambda}} - \boldsymbol{\lambda} \|_{\infty}$ \\
  \hline
  Ramesses & \ref{fig:CleanResults}(a) & $457,616$ & $9.52 \times 10^{-8}$ \\
  Raptor & \ref{fig:CleanResults}(b) & $1,666,700$ & $1.98 \times 10^{-8}$ \\
  Aquarius & \ref{fig:IncompleteDataResults}(a) & $176,771$ & $3.46 \times 10^{-7}$ \\
  Horse & \ref{fig:IncompleteDataResults}(a) & $149,459$ & $3.47 \times 10^{-7}$ \\
  \hline
\end{tabular}
\end{table}

From the statistics, it can be easily found that our quasi-solution provides very accurate results on both the clean
point cloud and the real data. The numerical solver for computing the exact solution runs out of memory on the two
examples -- Momento and Neptune in Fig.\ref{fig:CleanResults}. Thus, the errors cannot be evaluated and shown here.

\subsection{Noisy data}
In the following tests, we verify the robustness of our approach on input with noises at different levels. For a given
point set $\mathcal{P}$ with normal vectors $\mathcal{N}$, if the diagonal length of its bounding box is $d$, a new
point set with $\delta \%$ Gaussian noises is obtained as follows:
\begin{itemize}
\item $n_{\mathcal{G}}=\ulcorner \frac{\delta}{100} n_{\mathcal{P}} \urcorner$ points are selected from $\mathcal{P}$
into a sub-set $\mathcal{G}$ with $n_{\mathcal{P}}$ denoting the number of points in $\mathcal{P}$;

\item Randomly generate a set of scale with Gaussian distribution, $\mathcal{D}_G=\{d_i\}, \, i=1,2,\ldots,n_G$, with
$d_i \in [0, \delta d / 1000]$;

\item Impose the noises onto the points in $\mathcal{G}$ by $\mathbf{p}'_j=\mathbf{p}_j + d_j \mathbf{n}_j$ for all
$\mathbf{p}_j \in \mathcal{G}$.
\end{itemize}
Normal vectors on a set of noisy points are re-generated by the orientation-aware \textit{Principal Component Analysis}
(PCA) with $p$-nearest neighbors -- here $p=6$ is used in all tests.

\begin{figure}
\includegraphics[width=\linewidth]{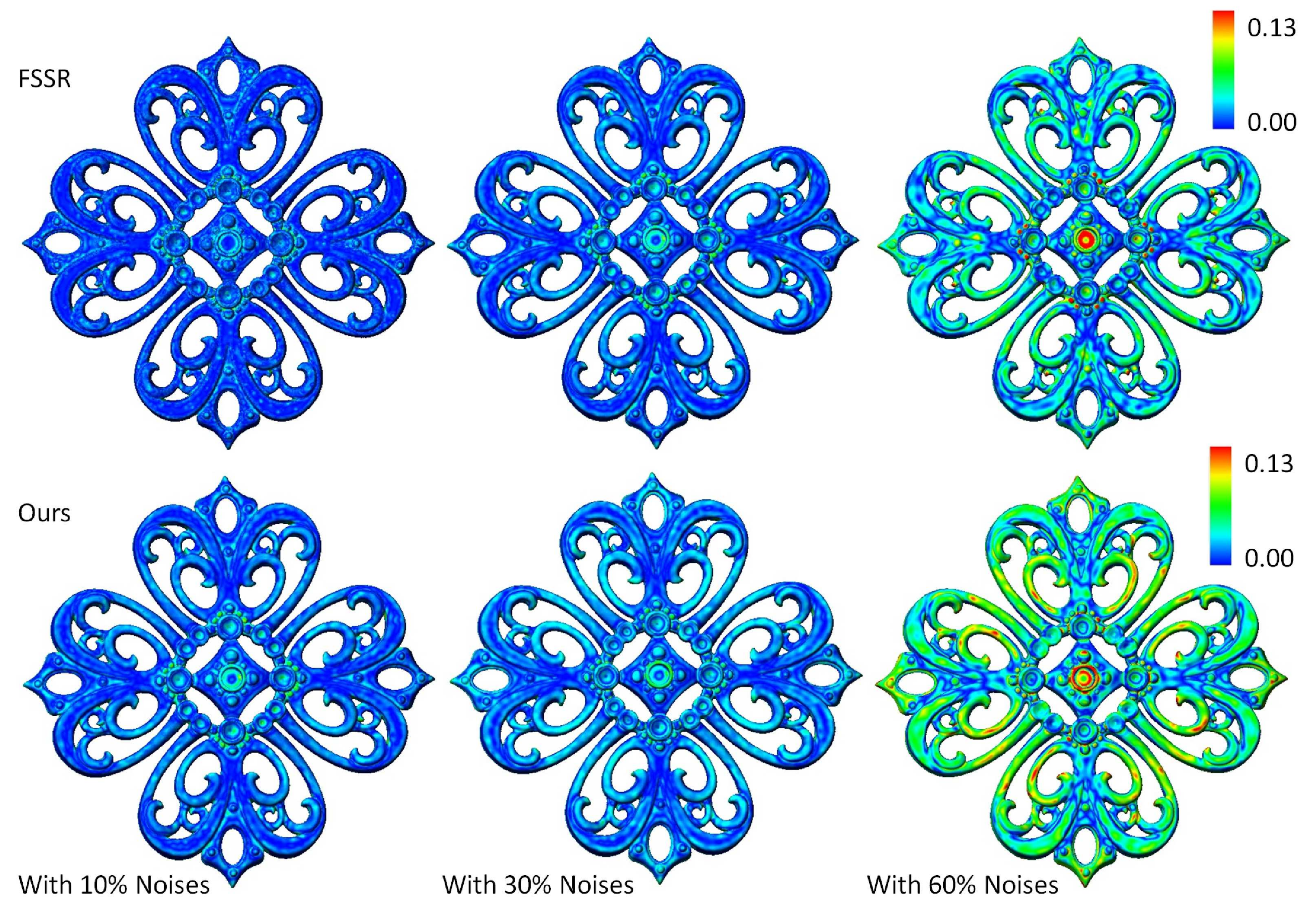}
\caption{Color maps to visualize the forward-distance based errors on the results generated by FSSR (top row) and our
method (bottom row).}\label{fig:NoisyExampleResultsShapeErr}\vspace{-10pt}
\end{figure}

\begin{table}[t]
\small \caption{Error Statistics of Reconstruction on Noisy Input}\label{tab:StatisticsOnNoisyData} 
\centering
\begin{tabular}{|c|c|c|c|c|c|}
  \hline
  \multicolumn{6}{|c|}{FSSR on the Filigree Model (Fig.\ref{fig:NoisyExampleResults})} \\
  \hline
  \textbf{Noisy} & \multicolumn{2}{c|}{\textbf{Forward Dist.}} & \multicolumn{2}{c|}{\textbf{Backward Dist.}} & \\
  \cline{2-5}
  \textbf{Level} & \textbf{Max.} & \textbf{Ave.} & \textbf{Max.} & \textbf{Ave.} & \textbf{Scale} \\
  \hline
  10\% & $.00360$  & $.000295$ & $.0175$  &  $.00315$ & $.00375$ \\
  30\% & $.00800$  &  $.00600$ &  $\mathbf{.0325}$  &  $.0014$ &  $.00670$ \\
  60\% & $.0115$  &  $.00145$ &  $\mathbf{.0300}$  &  $.00170$ &  $.0112$ \\
  \hline
  \hline
  \multicolumn{6}{|c|}{Our Method on the Filigree Model (Fig.\ref{fig:NoisyExampleResults})} \\
  \hline
  \textbf{Noisy} & \multicolumn{2}{c|}{\textbf{Forward Dist.}} & \multicolumn{2}{c|}{\textbf{Backward Dist.}} & \\
  \cline{2-5}
  \textbf{Level} & \textbf{Max.} & \textbf{Ave.} & \textbf{Max.} & \textbf{Ave.} & $s$ \\
  \hline
  10\% & $.00405$ &  $.000255$ & $.0115$  &  $.000275$ & $1.9$  \\
  30\% & $.00700$  &  $.000800$  & $.00700$  &  $.000800$  & $2.7$  \\
  60\% & $.0135$   &  $.00210$  & $.0150$ &  $.00220$  & $3.5$ \\
  \hline
\end{tabular}
\begin{flushleft}
$^{*}$The errors are measured by the Metro tool~\cite{Cignoni:1998}.
\end{flushleft}\vspace{-10pt}
\end{table}

We reconstruct mesh surfaces from a filigree model with $30\%$ and $60\%$ Gaussian noises by different methods,
including Screen-Poisson, FSSR and ours (see Fig.\ref{fig:NoisyExampleResults}). The noise-free set of points are
sampled from a mesh model so that we can compare the results of reconstruction with the original mesh to evaluate the
shape approximate errors generated by different method. Screen-Poisson reconstruction does not perform well on noisy
model. As shown in Fig.\ref{fig:NoisyExampleResults}, models with incorrect topology are generated. FSSR and ours can
still reconstruct `correct' models even after embedding $60\%$ Gaussian noises. We then compare these two methods in
terms of shape approximation error by the Metro tool~\cite{Cignoni:1998} (see Table \ref{tab:StatisticsOnNoisyData}).
In the measurements based on forward distances from ground-truth to the reconstruction, FSSR has slightly smaller
errors (see also the visualization in Fig.\ref{fig:NoisyExampleResultsShapeErr}). In the errors based on backward
distances (i.e., from reconstruction to ground-truth), our method outperforms FSSR. This is because FSSR generates some
interior isolated regions (i.e., topological errors) but our method does not -- see the zoom-view in
Fig.\ref{fig:NoisyExampleResults}. Moreover, our method is $17.5\times$ and $36.4\times$ faster than FSSR on the $30\%$
and $60\%$ noisy models respectively.

\subsection{Limitation}
\begin{figure}\centering
\includegraphics[width=.9\linewidth]{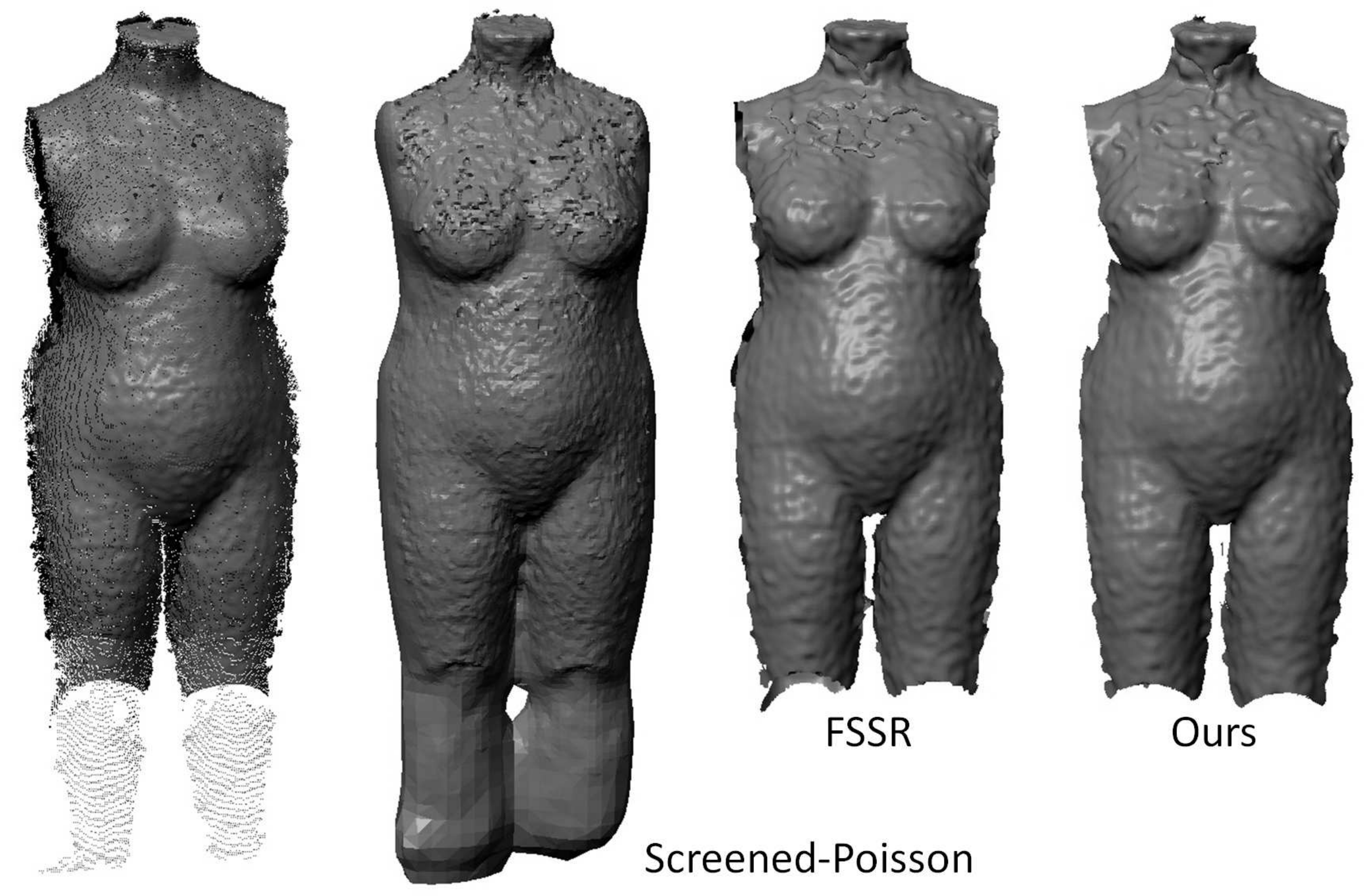}
\caption{For an input point set (left most) with low quality from Kinect, all methods generate poor
results.}\label{fig:Limitations}\vspace{-10pt}
\end{figure}

The limitations of our approach are mainly caused by the nature of locally compact support of kernel functions. As a
result, we share the following common limitations as the FSSR method.
\begin{itemize}
\item Near the boundary of regions with function-value defined, some small fragments isolated from the main
reconstruction could be formed by the numerical oscillation. Such isolated fragments must be removed by the
post-processing step taken on the mesh surface after polygonization.

\item Although reconstruction with high quality can be found at the example shown in
Fig.\ref{fig:FourSyntheticScansResult}, misaligned multiple scan could lead to multi-layers of points, therefore also
have multiple surface layers produced at those `overlapped' regions.
\end{itemize}
Caused by these limitations, when a set of points with low quality (e.g., the set obtained from two Kinect sensors as
shown in Fig.\ref{fig:Limitations}) are used as input, neither FSSR nor ours can obtain water-tight surface as
generated by Screen-Poisson reconstruction.

\section{Conclusion}
In this paper, we present a novel surface reconstruction method based on computing an approximate solution of
HRBF-based implicit surface fitting. The approximate solution is formulated as a weighted sum of compactly supported
basis functions centered at input data points equipped with normal vectors (i.e., we provide a closed-form solution
without any global operation). The implicit function for surface reconstruction can be efficiently and robustly
evaluated. The error-bound between our approximate solution and the exact solution has been derived, which can be
guaranteed as long as the maximum number of points covered in the supports of RBF kernels is capped by a fixed number.
Moreover, to strengthen the performance of our approach on input with high non-uniformity, a center selection algorithm
has also been introduced. Experimental results have shown the performance of our approach by comparing to the
state-of-the-arts.

No global operation needs to be applied during the surface reconstruction of our approach. As a result, it is easy to
extend our implementation to run in the out-of-core manner or on a distributed PC-cluster. We would like to further
investigate the strength of our method in this aspect in our future work, which can make it possible to realize the
on-site reconstruction of large-scale 3D models (e.g., outdoor scenes like city scale). Many robotics and virtual
reality applications could benefit from this work.


%

%

\ifCLASSOPTIONcompsoc
  \section*{Acknowledgments}
\else
  \section*{Acknowledgment}
\fi
The authors would like to thank the support provided by the HKSAR RGC General Research Fund (CUHK/14207414) and the
Natural Science Foundation of China (Ref. No.: 61173119).

\ifCLASSOPTIONcaptionsoff
  \newpage
\fi



\bibliographystyle{IEEEtran}
\bibliography{IEEEabrv,TVCGHRBFSurfRecon}

%


\vfill


\end{document}